\documentclass[%
reprint,
amsmath,amssymb,
aps,
pra,
]{revtex4-2}

\usepackage{graphicx}
\usepackage{dcolumn}
\usepackage{bm}

\begin{document}


\title{
Macroscopic Single-Qubit Operation for Coherent Photons
}

\author{Shinichi Saito}
 \email{shinichi.saito.qt@hitachi.com}
\affiliation{Center for Exploratory Research Laboratory, Research \& Development Group, Hitachi, Ltd. Tokyo 185-8601, Japan.}

\date{\today}

\begin{abstract}
Polarisation is described by an $SU(2)$ wavefunction due to macroscopic coherence of photons emitted from a ubiquitous laser source, and thus, a laser pulse is expected to behave as a macroscopic quantum bit (qubit), i.e., a qubit realised by a macroscopic number of photons.
Here, we show that an arbitrary single-qubit operation can be carried out for such a macroscopic qubit by employing optical modulators, together with standard optical plates, in a computer-controlled fibre-optic configuration. 
We named the device as a Poincar\'e rotator, which allows a dynamic control over a polarisation state by executing an arbitrary amount of rotations on the Poincar\'e sphere. 
The Poincar\'e rotator works as an arbitrary $SU(2)$ operator in a Lie group, by combining a $U(1)$ operation to change the phase and another $U(1)$ operation to change the amplitude of the wavefunction.
We have realised various polarisation states, such as $4 \times 4=16$, $8 \times 8=64$, and $10 \times 10=100$ distinguishable states on the sphere.
As a locus of the realised polarisation states on the sphere, we have successfully drawn the molecular structure of Buckminsterfullerene (C$_{60}$) and the coastline of the earth.
\end{abstract}


\maketitle

\section{Introduction}

The notion to realise macroscopic quantum phenomena, such as Macroscopic Quantum Tunnelling (MQT)\cite{Caldeira81,Leggett95,Coleman77,Coleman88} and Macroscopic Quantum Coherence (MQC) \cite{Leggett85}, is of fundamental importance to understand the nature to exhibit apparent differences between a classical macroscopic world and a quantum-mechanical microscopic world.
MQT would also be important to elucidate the birth of the universe \cite{Coleman77,Coleman88,Weinberg05}, while MQC is expected for applications in quantum computing \cite{Nielsen00} using superconducting quantum bits (qubits) \cite{Nakamura99,Koch07,Schreier08,Arute19}.
For observing MQT and MQC in experiments, it is essential to reduce dissipation of energies to environments \cite{Caldeira81}, which prevents quantum tunnelling, as shown by an imaginary-time path-integral calculation along an instanton saddle point \cite{Caldeira81,Leggett85,Leggett95,Coleman77,Coleman88}. 
They will also contribute to elucidating a paradox of Schr\"odinger's cat \cite{Caldeira81,Leggett85,Leggett95,Coleman77,Coleman88,Takagi02,Jaeger14,Frowis18}.

Aside from MQT and MQC, there is a different class of macroscopic quantum phenomena, such as superconductivity \cite{Bardeen57,Anderson58,Schrieffer71}, superfluidity \cite{Matsubara56,Ginzburg50}, Bose-Einstein Condensation (BEC) \cite{Cornell02}, and a laser \cite{Max99,Jackson99,Yariv97,Gil16,Goldstein11,Hecht17,Pedrotti07}, where a macroscopic number of elementary particles occupy the same quantum level to achieve phase coherence in a system.
A macroscopic quantum phenomenon of this kind is classified as the type-I, where only 1 degree or a few degrees of freedom describe an entire system quantum-mechanically, while MQT and MQC are classified as the type-II, where so many degrees of freedom are involved, such that quantum-mechanical behaviours are much more difficult to observe \cite{Takagi02,Jaeger14,Frowis18}.
For example, in a case of superconductivity  \cite{Bardeen57,Anderson58,Schrieffer71,Bogoljubov58,Abrikosov75,Fetter03,Kittel04,Nagaosa99,Altland10} as for a type-I macroscopic quantum phenomenon, conducting electrons, which are fermions, form Cooper pairs below a critical temperature to exhibit a BEC-like condensation similar to bosons, such that the entire system is described by a sole wavefunction, which satisfies the Ginzburg-Landau (GL) equation \cite{Ginzburg50}.
The phenomenological GL equation was microscopically justified by a weak coupling Bardeen-Cooper-Schriefer (BCS) theory \cite{Bardeen57,Schrieffer71}, which is continuously connected to a strong coupling theory of BEC \cite{Matsubara56}, known as a BCS-BEC crossover \cite{Nozieres85,Micnas90,Suzuki99,Saito01,Saito19}. 
Consequently, it is well understood as a paradigm that superconductivity is a macroscopic quantum phenomenon, whose property is described by a macroscopic wavefunction, as a consequence of the broken gauge symmetry of the unitary group of the order 1 ($U(1)$) \cite{Nambu59,Anderson58,Goldstone62,Higgs64,Schrieffer71}.
Cooper pairs also have internal degrees of freedom to form singlets or triplets, with and without orbital angular momentum, reflecting many-body interaction between pairs and symmetries of materials \cite{Kittel04,Nagaosa99,Altland10,Leggett04}.

We have revisited an issue of a broken symmetry for a laser \cite{Saito20a,Saito20b,Saito20c,Saito20d,Saito20e,Saito21f,Saito22g}, and found that the onset of lasing beyond the threshold of pumping is in fact related to be a broken rotational symmetry of polarisation \cite{Saito20d} due to the selection of the mode in the cavity with the lowest propagation loss for continuous processes of stimulated emissions \cite{Yariv97,Parker05,Chuang09}.
The broken symmetry is described by the special unitary group of the order 2 ($SU(2)$) in the Lie group \cite{Stubhaug02,Fulton04,Hall03,Pfeifer03,Dirac30,Georgi99}, since the polarisation is  a macroscopic manifestation of spin of photons \cite{Saito20a}.
We found a decent similarity between the BCS theory \cite{Bardeen57} and an $SU(2)$ theory for a confined mode in a graded-index (GRIN) fibre \cite{Saito20d}, and the energy gap in a superconductor corresponds to the order parameters for photons, which are Stokes parameters, obtained as spin expectation values \cite{Saito20a,Saito20d}.
Due to the BEC-like nature of lasing, the lasing mode is simply characterised by a wavefunction of photons \cite{Saito20d} to represent the mode in the waveguide or the fibre of the laser, which satisfies the Helmholtz equation, derived from the Maxwell equations  \cite{Max99,Jackson99,Yariv97,Gil16,Goldstein11,Hecht17,Pedrotti07}.
Consequently, macroscopic number of coherent photons emitted from a laser are simply represented by an $SU(2)$ wavefunction for polarisation, and the simple quantum-mechanical expectation values of spin components, calculated by the wavefunction \cite{Jones41,Fano54,Baym69,Sakurai67,Sakurai14,Saito20d}, are indeed Stokes parameters \cite{Stokes51} represented in the Poincar\'e sphere \cite{Poincare92}.
It is astonishing to consider why Stokes could arrive right formulas \cite{Stokes51} even before the discovery of the quantum mechanics \cite{Plank00,Einstein05,Bohr13,Dirac28} and why Poincar\'e could propose an intuitive representation in the sphere \cite{Poincare92} before the establishment of mathematics of the representation theory of Lie groups and Lie algebras \cite{Stubhaug02,Fulton04,Hall03,Pfeifer03,Dirac30,Georgi99}.
From a modern perspective, we have all necessary theoretical frameworks in physics and mathematics, including quantum many-body field theories \cite{Sakurai67,Abrikosov75,Fetter03,Nagaosa99,Weinberg05,Wen04,Altland10,Fox06}, and experimentally, we have a wide range of choices on lasers and optical components \cite{Max99,Jackson99,Yariv97,Parker05,Chuang09,Grynberg10,Gil16,Goldstein11,Hecht17,Pedrotti07}.

These theoretical backgrounds suggest that we can use the $SU(2)$ degrees of freedom in coherent photons emitted from a laser as for a macroscopic qubit.
In fact, the Poincar\'e sphere \cite{Poincare92,Jones41,Fano54} is equivalent to the Bloch sphere \cite{Arecchi72,Narducci74,Baym69,Sakurai14}, except for the fact that the radius of  the Poincar\'e sphere is proportional to the number of photons \cite{Saito20a,Saito20d}, while the radius of the Bloch sphere is normalised to be 1 for a single electron.
For photons, there is no difficulty to occupy the same polarisation state unlike fermionic electrons, since photons are bosons, such that the polarisation state is represented by an $SU(2)$ wavefunction, even if the number of photons is large. 
Therefore, if we make a pulse stream of coherent photons, each pulse can be represented by an $SU(2)$ state, whose average spin value is shown on the Poincar\'e sphere.
As discussed above, this macroscopic character of coherent photons is simply coming from the type-I nature of a macroscopic quantum phenomenon, such that macroscopic number of photons are degenerate to be settled in the state with the lowest propagation loss upon lasing.
In this context, the pulse stream of coherent photons is not representing a type-II macroscopic quantum phenomenon, nor the pulse is not representing a Schr\"odinger's cat state.
Moreover, we are not intending to demonstrate a two-qubit operation \cite{Nielsen00} in a macroscopic scale, which is far beyond the scope of this paper.
It is not clear at present whether a two-qubit operation can be achievable using coherent photons, since no-cloning theorem prevents to copy an entangled state by a unitary operation without measuring the state \cite{Wootters82,Dieks82}.
In that sense, we are not intending to apply a macroscopic qubit of coherent photons for immediate applications in quantum computing \cite{Nielsen00,Nakamura99,Koch07,Schreier08,Arute19,Bruzewicz19,Pino21,
OBrien03,Peruzzo14,Silverstone16,Takeda17,Lee20,Xue21,Preskill18}.

Nevertheless, an $SU(2)$ state can accommodate much more analogue information than a binary digital information, such that macroscopic qubits will be useful for transmitting big data for communications.
In fact, an $SU(2)$ state is characterised by the 2-dimensional sphere of the unit length, $S^{2}=\{ {\bf x} \in  \mathbb{R}^{2} | |{\bf x}|=1 \}$, which is topologically isomorphic to the Poincar\'e sphere, such that it is described by 2 continuous valuables of the polar angle ($\theta$) and the azimuthal angle ($\phi$) \cite{Max99,Jackson99,Yariv97,Parker05,Chuang09,Grynberg10,Gil16,Goldstein11,Hecht17,Pedrotti07,
Saito20a,Saito22g}, while the binary information is based on the cyclic group of the order 2, $\mathbb{Z}_2=\{ 0, 1 \}$, which is isomorphic to the 0-dimensional sphere of the unit length, $S^{0}=\{ {\bf x} \in  \mathbb{R} | |{\bf x}|=1 \}=\{ -1, 1 \}$ \cite{Stubhaug02,Fulton04,Hall03,Pfeifer03}.
A full amount of utilisations on continuous variables of both $\theta$ and $\phi$ as macroscopic qubits have a potential to expand the communication bandwidths, significantly.

So far, only a limited amount of the polarisation capabilities are being used for applications in high-speed optical communications \cite{Kikuchi16,Debnath18,Zhang21} among various advanced communication formats like Quadrature-Amplitude-Modulation (QAM) and Pulse-Amplitude-Modulation (PAM).
For example, polarisation modulators by converting from a Transverse-Electric (TE) mode to a Transverse-Magnetic (TM) mode, or {\it vice versa}, are employed using LiNbO$_3$ (LN) \cite{Thaniyavarn86} and AlGaAs \cite{Bull04} with a 3 dB bandwidth exceeding 40 GHz, however, these modulators are controlling the polarisation only along the circles of latitude (parallels) or circles of longitude (meridians), individually.
It is not yet introduced to control both parallels and meridians, simultaneously, for scanning the full Poincar\'e sphere.
Another examples are found in Dual-Polarisation Quadrature-Phase-Shift-Keying (DP-QPSK) using Si photonic platforms \cite{Goi14,Doerr15}, which increases a bandwidth by a factor of 2 due to the orthogonal duality in polarisation, but full dynamic multiplexing on polarisation degrees of freedom together with QPSK will significantly increase the bandwidth with factors of $8\times8=64$, for example, depending on the multiplexing over $\theta$ and $\phi$ for the future.
Novel technologies using 2-dimensional materials \cite{Tiu22} and non-linear metamaterials \cite{Nicholls17} are also developed for polarisation controls, and applications to polarisation qubits using single photons are also  demonstrated by on-chip polarisation control to realise arbitrary wave plates \cite{Heilmann14}.
More traditionally, chiral materials like liquid crystals or quartz are used for displays and passive controls  \cite{Gil16,Goldstein11,Hecht17,Pedrotti07}, but the operation speed is limited to be less than $\sim$ 1 kHz.
Regardless of these advanced technologies, full dynamic controls over the Poincar\'e sphere at high-speed still remain challenging.

Previously, we have proposed a passive Poincar\'e rotator, which allows an arbitrary amount of rotation simply by combining half-wave-plates (HWPs) and quarter-wave plates (QWPs) \cite{Saito22g}.
For realising a genuine rotator of $U(1)$ by a physical rotation of a HWP, rather than a pseudo rotator \cite{Gil16,Goldstein11}, it was critical to introduce another HWP \cite{Saito22g}.
By using 2 HWPs, we could convert the pseudo rotator to the genuine rotator of a mirror reflection simply by rotating physically \cite{Saito22g}.
Then, a phase-shifter could be constructed by inserting 2 QWPs before and after the genuine rotator \cite{Saito22g}.
Consequently, a passive Poincar\'e rotator is a convenient tool to realise an arbitrary $SU(2)$ quantum operation to a macroscopic wavefunction of the spin state of coherent photons to describe polarisation \cite{Saito20a,Saito22g}.
We think it is not trivial to expect that we can control a quantum state of a macroscopic wavefunction.
For example, in the case of superconductivity, the $U(1)$ gauge symmetry is broken, and the Nambu-Anderson-Goldstone boson \cite{Nambu59,Anderson58,Goldstone62,Higgs64,Schrieffer71} guarantees to change the $U(1)$ at zero-energy due to a Mexican hat-shape free energy to have saddle points along a rotational direction \cite{Abrikosov75,Fetter03,Nagaosa99,Wen04,Altland10}.
However, electrons have charge, such that the collective mode of the condensate induces a plasma oscillation and the excitation energy becomes massive.
Therefore, the continuous change of the $U(1)$ degree of freedom is not easy to observe, and we need a Josephson junction, for example, to see the impacts of changes in the $U(1)$ phase.
On the other hand, photons do not have charge, and a similar argument of the BCS theory holds to ensure to change the $SU(2)$ wavefunctions at zero-energy \cite{Saito20d}.
In fact, it is easy to change the phases of $U(2)$ in the wavefunction for 2 orbital degrees of freedom into the phase difference in spin degrees of freedom, simply by inserting the ray of coherent photons into the wave-plates, which breaks the rotational symmetry along slow and fast axes \cite{Saito20a,Saito22g}.

Here, we extend the idea of a Poincar\'e rotator to be an active device.
An active Poincar\'e rotator allows us to realise a time-dependent $SU(2)$ operator, which acts on a macroscopic wavefunction of coherent photons to realise an arbitrary rotation to cover the entire polarisation state on the Poincar\'e sphere.
This corresponds to realise an arbitrary single-qubit operation for the pulse stream of the coherent ray from a standard laser. 

\section{Theory}

We will briefly explain about the theory how to construct an active Poincar\'e rotator for manipulating polarisation state in an $SU(2)$ theory \cite{Jones41,Fano54,Baym69,Sakurai67,Sakurai14,Saito20d,Saito22g}.
The Poincar\'e rotator is composed of the rotation operators $\hat{\mathcal{R}}_i (\delta \phi)$ along the axes of $i=1,2,3$ of Stokes parameters, $(S_1,S_2,S_3)$, on the Poincar\'e sphere for the rotation angle of $\delta \phi$.
We obtain $\hat{\mathcal{R}}_i (\delta \phi)$ from a group element of an $SU(2)$ Lie group by an exponential map \cite{Stubhaug02,Fulton04,Hall03,Pfeifer03,Dirac30,Georgi99} as
\begin{eqnarray}
\hat{\mathcal{R}}_i (\delta \phi)
&\equiv&
\hat{\mathcal{D}} ({\bf \hat{n}}_i,\delta \phi) 
=\exp 
\left (
-i 
\hat{\bm \sigma} \cdot {\bf \hat{n}}_i
\left (
\frac{\delta \phi}{2}
\right)
\right), 
\end{eqnarray}
where ${\bf \hat{n}}_i$ is the unit vector along $S_i$ axis, and the spin operator $\hat{\bm \sigma}=(\sigma_3, \sigma_1, \sigma_2)$ is given by Pauli matrices are given by 
\begin{eqnarray}
\sigma_1=
\left(
  \begin{array}{cc}
0 & 1 \\
1 & 0
  \end{array}
\right),
\sigma_2=
\left(
  \begin{array}{cc}
0 & -i \\
i & 0
  \end{array}
\right) , 
\sigma_3=
\left(
  \begin{array}{cc}
1 & 0 \\
0 & -1
  \end{array}
\right), \nonumber \\
\end{eqnarray}
which form a basis of a Lie algebra of $\mathfrak{su}(2)$  \cite{Baym69,Sakurai14,Fulton04,Hall03,Pfeifer03,Georgi99}.
Here, we have used a basis for the polarisation to diagonalise horizontal (H) and vertical (V) states, whose principal quantisation axis is assigned to be $S_1$.
If we would like to use another basis, e.g., with left (L) and right (R) states can be used simply by setting $\hat{\bm \sigma}=(\sigma_1, \sigma_2, \sigma_3)$ \cite{Saito20a,Saito20c,Saito22g}.
But, it will be convenient to use a HV basis rather than LR states for an active Poincar\'e rotator, since a polarisation splitter is widely available for HV states. 
It is also useful to be aware of the formula \cite{Baym69,Sakurai14,Saito20a,Saito22g}
\begin{eqnarray}
\hat{\mathcal{R}}_i (\delta \phi)
=
{\bf 1}
\cos
\left (
\frac{\delta \phi}{2}
\right)
-i 
\hat{\bm \sigma} \cdot {\bf \hat{n}}_i
\sin
\left (
\frac{\delta \phi}{2}
\right)
, 
\end{eqnarray}
where ${\bf 1}$ is the identity matrix of $2 \times 2$.

For the construction of an passive Poincar\'e rotator, the crucial step was to make a rotator along the $S_3$ axis by combining 2 HWPs to allow a $U(1)$ rotation to form a rotational group \cite{Saito22g}.
Then, it was straightforward to make the phase-shifter, which rotates a polarisation state along the $S_1$ axis simply by using another 2 QWPs to change the rotation axes from $S_3$ to $S_1$ and bring them back to the original coordinate \cite{Saito22g}.
For the present active Poincar\'e rotator, the building process becomes opposite, namely, we will consider to make a $U(1)$ phase-shifter, first, and then, apply it to make a rotator to realise a full $SU(2)$ rotation.
The reason is that we can utilise an optical modulator to allow the phase-shift for the wave propagating in the waveguide, such that we can easily control the phase-shift between H and V states.

\subsection{Phase-shifter}

For the physical realisation of the phase-shifter, we need to build a device to operate as 
\begin{eqnarray}
\hat{\mathcal{R}}_1 (\delta \phi)
&\equiv&
\hat{\mathcal{D}} ({\bf \hat{n}}_1,\delta \phi) 
=\exp 
\left (
-i 
{\sigma}_3
\left (
\frac{\delta \phi}{2}
\right)
\right), 
\end{eqnarray}
for polarisation state.
This could be achieved first by splitting a ray from a laser between $|{\rm H} \rangle$ and $|{\rm V} \rangle$ by a polarisation-beam-splitter (PBS), such that the splitter works as a projection operator
\begin{eqnarray}
\hat{\mathcal{P}}_{HV} 
=
|{\rm H} \rangle \langle {\rm H} |
+
|{\rm V} \rangle \langle {\rm V} |
, 
\end{eqnarray}
and then, phases-shifts of ${\rm e}^{-i  \delta \phi/2}$ and ${\rm e}^{i  \delta \phi/2}$ were applied for H and V states, respectively.
Two optical modulators were employed to guarantee the same optical lengths in a push-pull configuration to give the opposite phases.
After the phase-shift, the rays for $|{\rm H} \rangle$ and $|{\rm V} \rangle$ states are combined by a polarisation-beam-combiner (PBC), which is actually the same optical component with the PBS, but inputs and outputs are opposite each other, due to the time-reversal symmetry.
Finally, we realise
\begin{eqnarray}
\hat{\mathcal{R}}_1 (\delta \phi)
=
\left(
\begin{array}{cc}
 {\rm e}^{-i  \delta \phi/2} & 0 \\
0 & {\rm e}^{i  \delta \phi/2}
  \end{array}
\right), 
\end{eqnarray}
where $\delta \phi$ is determined by voltages applied to optical modulators.
In the actual experiments, we have used just 1 arm for the phase-shift to simplify the active electronic control to focus on 1 modulator in this study, which corresponds to make a phase-shifter of 
\begin{eqnarray}
\hat{\mathcal{R}}_1^{\prime} (\delta \phi)
=
\left(
\begin{array}{cc}
1 & 0 \\
0 & {\rm e}^{i  \delta \phi}
  \end{array}
\right), 
\end{eqnarray}
which is equivalent to the original $\hat{\mathcal{R}}_1 (\delta \phi)$, except for the global phase factor.
As far as we are dealing with individual bits, the overall phase factor does not play any role at all, but if we would like to compare the phase difference among the bits by an interference experiment, a push-pull operation would be required, which also save the applied voltages to modulators by a factor of 2.

\begin{figure}[h]
\begin{center}
\includegraphics[width=8cm]{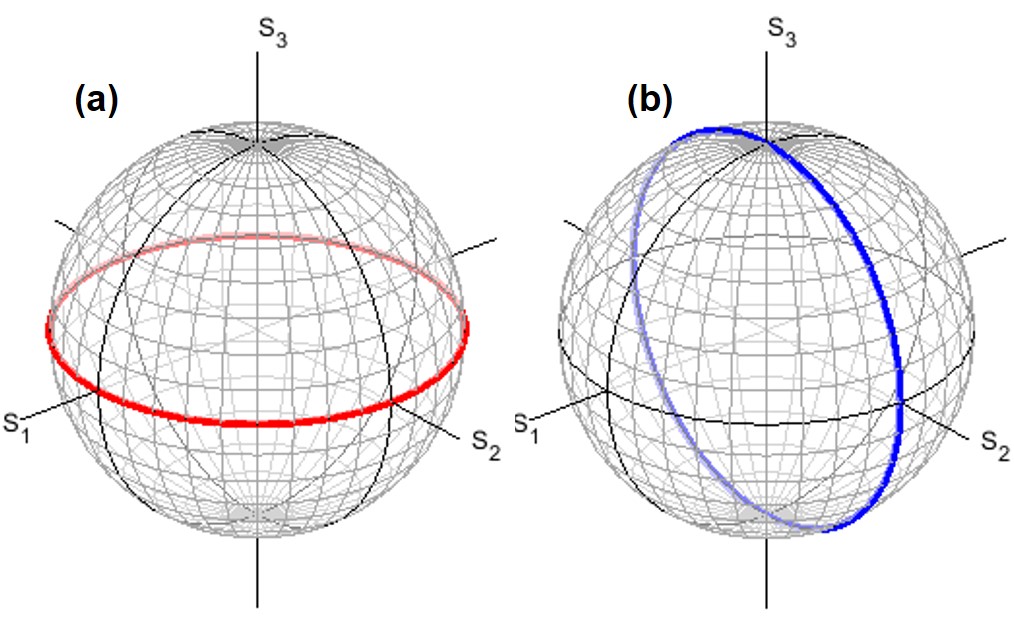}
\caption{
Expected polarisation states controlled by a proposed Poincar\'e rotator.
The input state was the diagonally polarised state, and rotation angles were changed continuously.
Trajectories of states controlled by (a) rotator and (b) phase-shifter configurations.
}
\end{center}
\end{figure}

An example of a calculated trajectory on polarisation states controlled by the phase-shifter is shown in Fig. 1 (b) for the input of the diagonally (D) polarised state.
As we increase the phase-shift, the states are converted from D to L, A (anti-diagonally polarised state), R, and D states, circulating along with the meridian.
The phase-shift of $\delta \phi$ simply becomes the amount of rotation in the Poincar\'e rotator.
Here, it is important to guarantee to include the identity operation of ${\bf 1}$ to ensure the polarisation state without changing the input state.
In the case of a pseudo rotator, realised by a physical rotation of HPW, this condition has not been met, such that pseudo rotators could not form a proper sub-group in $SU(2)$ \cite{Gil16,Goldstein11,Saito22g}.
On the other hand, in the proposed set-up, the identity operation could be realised by equalising the optical path lengths between 2 arms for H and V states to vanish the phase-shift of $\delta \phi=0$.



\subsection{Rotator}

Next, we consider to make a rotator out of the phase-shifter (Fig. 1 (a)).
First, we will rotate the polarisation state along the $S_2$ axis towards the anti-clock-wise direction for the rotation angle of $\pi/2$.
This was achieved by rotating a QWP to align its fast-axis (FA) to pointing the direction, which has an angle of $\pi/4$ from the horizontal direction in the anti-clock-wise direction.
We obtain the corresponding operation as 
\begin{eqnarray}
\hat{\mathcal{D}} ({\bf \hat{n}}_2,\pi/2) 
=
\frac{1}{\sqrt{2}}
\left(
\begin{array}{cc}
1 & -i \\
-i & 1
\end{array}
\right)
, 
\end{eqnarray}
which brings the L state (the north pole in our convention) to the H state, for example.
This corresponds to exchange the $S_3$ axis with the $S_1$ axis. 
Then, we will apply the phase-shifter operation of $\hat{\mathcal{R}}_1^{\prime} (\delta \phi)$, made of 2 optical modulators, for the amount of $\delta \phi$, which we wish to rotate along the $S_1$ axis as a phase-shifter.
Finally, we will bring the rotated axes back to the original coordinate by an opposite rotation using a QWP to align its FA to pointing the direction, which has an angle of $\pi/4$ from the horizontal direction in the clock-wise direction.
\begin{eqnarray}
\hat{\mathcal{D}} ({\bf \hat{n}}_2,-\pi/2) 
=
\frac{1}{\sqrt{2}}
\left(
\begin{array}{cc}
1 & i \\
i & 1
\end{array}
\right)
. 
\end{eqnarray}
The total process corresponds to the operation
\begin{eqnarray}
\hat{\mathcal{D}} ({\bf \hat{n}}_3,\delta \phi) 
&=&
\frac{1}{2}
\left(
\begin{array}{cc}
1 & i \\
i & 1
\end{array}
\right)
\left(
\begin{array}{cc}
 {\rm e}^{-i  \delta \phi/2} & 0 \\
0 & {\rm e}^{i  \delta \phi/2}
  \end{array}
\right)
\left(
\begin{array}{cc}
1 & -i \\
-i & 1
\end{array}
\right) 
\nonumber \\
&=&
\left(
\begin{array}{cc}
\cos(\delta \phi/2) & -\sin(\delta \phi/2)  \\
\sin(\delta \phi/2)  & \ \ \ \cos(\delta \phi/2) 
\end{array}
\right) 
\nonumber \\
&=&
\hat{\mathcal{R}}_3 (\delta \phi)
, 
\end{eqnarray}
which is exactly what we expected for the rotator, since the rotation on the Poincar\'e sphere corresponds to the twice of the rotation of the polarisation ellipse (${\it \Delta \Psi}$) in the real space for complex electric fields, such that we obtain $\delta \phi=2 {\it \Delta \Psi}$ \cite{Saito20a,Saito22g}.
As shown in Fig. 1 (a), we confirmed an expected trajectory for the input of the D state, rotated by the operation of $\hat{\mathcal{R}}_3 (\delta \phi)$.

\subsection{Poincar\'e rotator}

Having obtained both a rotator to operate as $\hat{\mathcal{R}}_3 (\delta \phi)$ and a phase-shifter to operate as $\hat{\mathcal{R}}_1 (\delta \phi)$, we can realise an arbitrary rotation by combining these 2 rotations.
For example, by using the H state as an input, we obtain
\begin{eqnarray}
 |\gamma, \delta \rangle
&=&
\hat{\mathcal{R}}_1 (\delta)
\hat{\mathcal{R}}_3 (\gamma)
|{\rm H} \rangle
\nonumber \\
&=&
\left(
\begin{array}{c}
 {\rm e}^{-i  \delta/2} \cos \alpha   \\
 {\rm e}^{+i  \delta/2} \sin \alpha  
\end{array}
\right) 
,
\end{eqnarray}
where $\gamma=2\alpha$ is the polar angle measured from $(1,0,0)$ in the Poincar\'e sphere, $\alpha$ is the auxiliary angle for complex electric fields, and $\delta$ is the relative phase between V and H states \cite{Saito20a,Saito22g}.
Therefore, the proposed Poincar\'e rotator corresponds to the operation of
\begin{eqnarray}
\hat{\mathcal{R}} (\gamma,\delta)
&=&
\hat{\mathcal{R}}_1 (\delta)
\hat{\mathcal{R}}_3 (\gamma)
\nonumber \\
&=&
\left(
\begin{array}{cc}
 {\rm e}^{-i  \delta/2} \cos (\gamma/2) &        - {\rm e}^{-i  \delta/2} \sin (\gamma/2)  \\
 {\rm e}^{+i  \delta/2} \sin (\gamma/2)  &   \ \ \ {\rm e}^{+i  \delta/2} \cos (\gamma/2)
\end{array}
\right) .
\nonumber \\
\end{eqnarray}

A care must be taken for the order of the rotations, since matrix operations depend on order of operations, which are coming from the non-Abelian nature of Lie algebra \cite{Fulton04,Hall03,Pfeifer03,Georgi99,Baym69,Sakurai14}.
We have constructed the Poincar\'e rotator, first by applying the rotator operation of $\hat{\mathcal{R}}_3 (\gamma)$, followed by the phase-shifter operation of $\hat{\mathcal{R}}_1 (\delta)$.
These operations are sequential and we cannot change the order of operation, after the assembly to set-up the order of operations.
Consequently, rotation planes are tilted, if the phase-shifter gives a finite change of the phases.
This is evident from Fig. 2 (a), where we have calculated trajectories of polarisation states for fixed discrete values of $\delta$.
At $\delta=0$, the trajectory lies in the $S_1$-$S_2$ plane (Fig. 1 (a)), while the constant value of $\delta$ tiles the original rotation axis of $S_3$ in $\hat{\mathcal{R}}_3 (\gamma)$ within the $S_3$-$S_2$ plane, such that the rotator operation corresponds to scan along meridians.
This is consistent with our choice of the HV basis, whose principal quantisation axis is along $S_1$, corresponding to the $\sigma_3$ operation as for the $z$-axis.
Therefore, the rotator operation of $\hat{\mathcal{R}}_3 (\gamma)$ is significantly affected by the following the phase-shift operation.

On the other hand, for the phase-shift operation of $\hat{\mathcal{R}}_1 (\delta)$, the rotator operation of $\hat{\mathcal{R}}_3 (\gamma)$ is completed before entering to the phase-shifter, such that the rotator merely changes the input state (Fig. 2 (b)).
Therefore, trajectories by the $\hat{\mathcal{R}}_1 (\delta)$ correspond to parallels, which are indeed parallel to the $S_2$-$S_3$ plane, similar to the $x$-$y$ plane in the right hand coordinate.
Please also note H and V states were not affected by a phase-shifter, since these states did not contain any contributions from the orthogonal states.
Namely, the H state cannot be changed by the phase-shift of the V state, since the H state is completely orthogonal to the V state, and {\it vice versa}, such that H and V states are stable against the operations of the phase-shift.
This is also consistent with our choice of the primary axis, such that H and V states correspond to north and south poles, respectively, in the HV basis. 

\begin{figure}[h]
\begin{center}
\includegraphics[width=8cm]{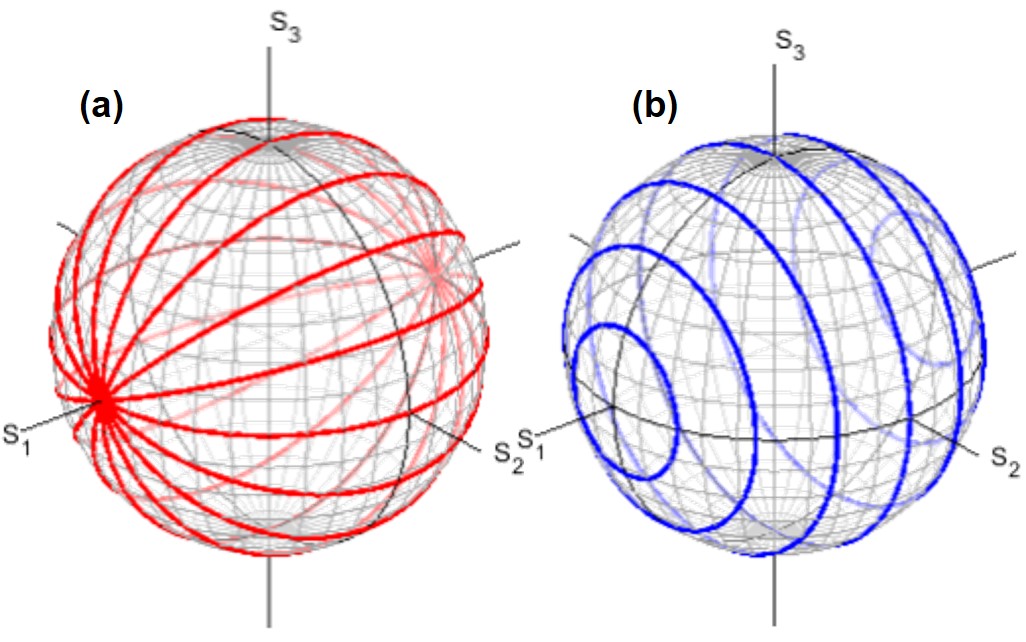}
\caption{
Impacts of successive rotations by the Poincar\'e rotator for the operation of $\hat{\mathcal{R}} (\gamma,\delta)$.
Polarisation states were calculated, assuming the input of the diagonally polarised state.
(a) continuous rotator operations for discrete phase-shifts of $\delta$ from $-\pi/2$ to $\pi/2$ at the $\pi/8$ step. 
(b) continuous phase-shifter operations for discrete rotations of $\gamma$ from $-\pi/2$ to $\pi/2$ at the $\pi/8$ step. 
}
\end{center}
\end{figure}

If one would like to use the $S_3$ axis as the primary axis, we can employ the LR basis instead of the HV basis.
In that case, we can make an arbitrary state as 
\begin{eqnarray}
 |\theta,\phi \rangle
&=&
\hat{\mathcal{R}}_3 (\phi)
\hat{\mathcal{R}}_2 (\theta)
|{\rm L} \rangle
\nonumber \\
&=&
\left(
\begin{array}{c}
 {\rm e}^{-i  \phi/2} \cos (\theta/2)  \\
 {\rm e}^{+i  \phi/2} \sin (\theta/2) 
\end{array}
\right),
\end{eqnarray}
where $\theta$ is the polar angle measured from $S_3$ and $\phi$ is the azimuthal angle measured from $S_1$ \cite{Saito20a,Saito22g}.
The choice of the basis is completely up to our preference, but it was experimentally convenient to use the HV basis for the alignments.

By combining the rotator and the phase-shifter, we can rotate the polarisation state in an arbitrary fashion.
In the HV basis, the rotator changes the amplitudes of HV components, while the phase-shifter changes the relative phase between H and V states, as evidenced in the form of $|\gamma, \delta \rangle$.
In the LR basis, the rotation along the $S_2$ axis corresponds to the amplitude control between L and R states, while the azimuthal phase control is achieved by the rotation along the $S_3$ axis.

\section{Experiments}

\subsection{Experimental set-up}

\begin{figure}[h]
\begin{center}
\includegraphics[width=8cm]{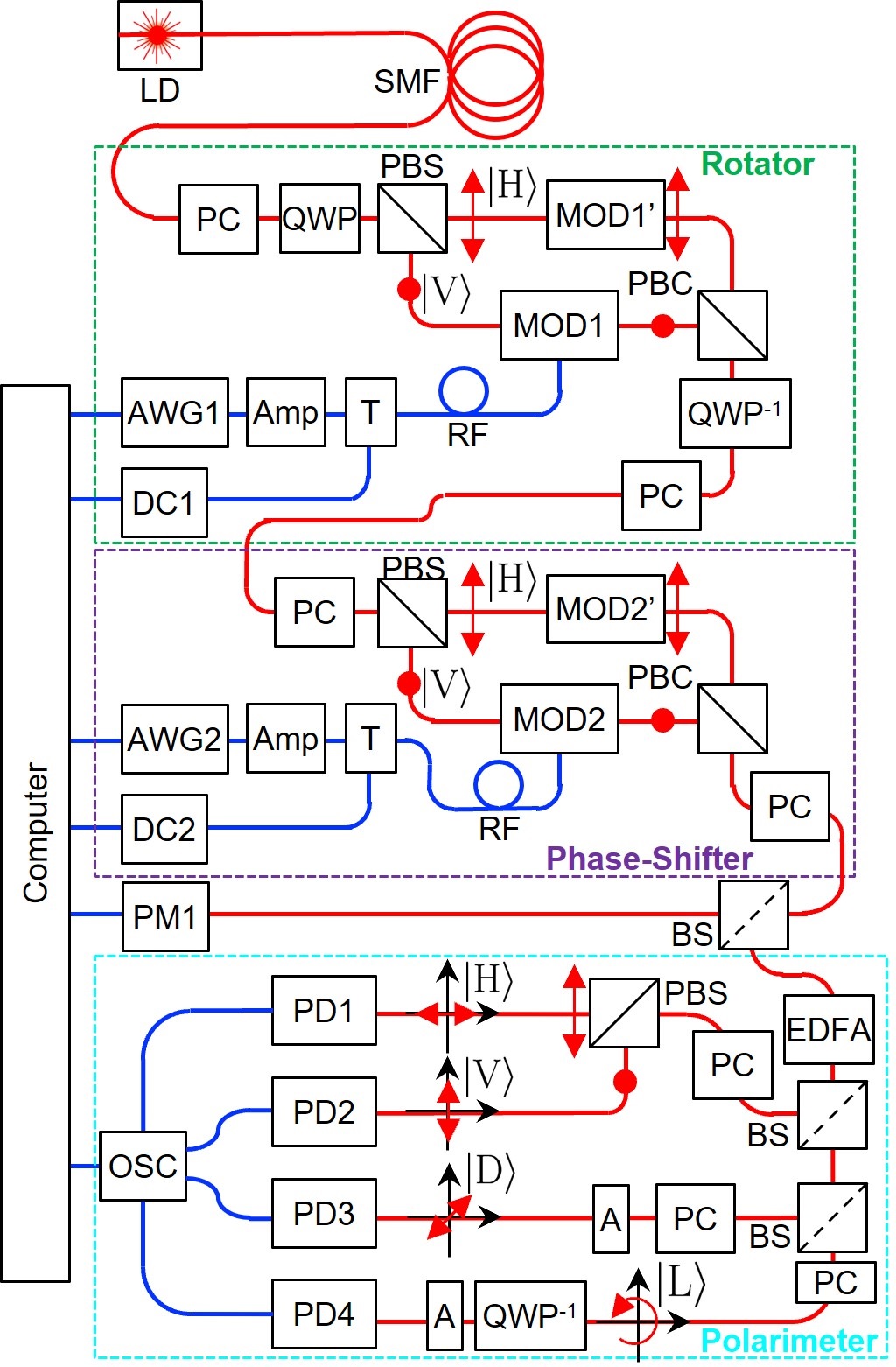}
\caption{
Poincar\'e rotator, which works as an arbitrary $SU(2)$ operator for polarisation states.
The input beam from a frequency locked laser at the wavelength of 1533 nm with the power of 1.8 mW was inserted to the rotator, followed by the phase-shifter.
The output beam was analysed by a standard polarimeter and a bespoke polarimeter for high-speed testing.
The active feedback system was constructed to compensate the DC drifts of optical modulators.
The optical modulators were electrically controlled by arbitrary-wave-form generators to enable active controls of polarisation states.
Abbreviations are as follows: 
LD: Laser Diode,  
SMF: Single Mode Fibre,  
PC: Polarisation controller,  
PBS: Polarisation Beam Splitter, 
PBC: PB Combiner,  
BC: Non-polarisation BS,  
QWP: Quarter-Wave-Plate,   
QWP$^{-1}$: Inverse QWP, 
A: A-filter, 
MOD: Modulator,   
PM: Polarimeter, 
RF: SMA coaxial cable, 
AWG: Arbitrary Waveform Generator,  
Amp: Amplifier, 
T: Bias-Tee,  
DC: DC power supply,   
EDFA: Er-Doped Fibre Amplifier, 
and  
OSC: Oscilloscope. 
}
\end{center}
\end{figure}

The experimental set-up for the proposed Poincar\'e rotator is schematically shown in Fig. 3.
A distributed-feedback (DFB) laser-diode (LD) with a frequency locking at the wavelength of 1533nm was used and the laser output is coupled to a single-mode-fibre (SMF).
A lot of passive polarisation-controllers (PCs), using stress-induced birefringence on SMFs, were used to make sure that the polarisation state is properly aligned during the propagation.
The Poincar\'e rotator is mainly comprised of 2 parts; (1) a rotator and (2) a phase-shifter, while we have also prepared (3) a polarimeter to observe the polarisation states controlled by the Poincar\'e rotator .

A rotator is the first component connected to the laser (the green part of Fig. 3 at the top).
The polarisation state of the laser output is adjusted to be $|{\rm D} \rangle$ as an input for a rotator.
In the rotator, the SMF is connected to a collimator lens (not shown) to propagate in a free space, where a QWP and a PBS were aligned, and after the polarisation splitting, each ray is connected to polarisation-maintaining fibre (PMF) assembled LN modulators (MODs), which were connected to collimator lenses to combine at a PBC, followed by the opposite rotation by a QWP (QWP$^{-1}$).
As explained above, the FA of the first QWP is aligned to rotate $\pi/4$ in the anti-clock-wise direction from the horizontal direction, while the FA of the second QWP (QWP$^{-1}$) is aligned to rotate $\pi/4$ in the clock-wise direction.
Finally, the combined ray is aligned to a collimator lens to a SMF.

Next step is to rotate the phase by a phase-shifter (the purple part of Fig. 3 at the middle).
Here, the set-up is the same as that for a rotator, except for the lack of 2 QWPs.
It is straightforward to see the role of the $U(1)$ rotation along the $S_1$ axis simply by changing the phase of $|{\rm V} \rangle$ state.
This is the simplest method experimentally to realise the operator of $\hat{\mathcal{R}}_1^{\prime} (\delta \phi)$, and the set-up is also ready to operate as $\hat{\mathcal{R}}_1 (\delta \phi)$ by a push-pull operation using MOD2' as well as MOD2.
This time, we have introduced MOD2' merely because of the equalisation of the optical path lengths during the phase-shift.

Both the rotator and the phase-shifter were connected to computer-controlled arbitrary-waveform-generators (AWGs), which were connected to electric amplifiers (Amps) and bias-tees (Ts).
The performance of the amplifier was shown in Supplementary Fig. 1, and it had the voltage gain of $\times 12.7$, which is equivalent to the power gain of 22.1 dB at 100 MHz, and the frequency bandwidth of the amplifier was from 5 MHz to 1GHz.
Our LN modulators were phase modulators and had a bandwidth of 10 GHz, but the modulation speed was limited by the bandwidth of 400 MHz due to the AWGs, used in our experiments.
Standard SMA (Sub-Miniature version-A) cables were used for the RF connections.
We also used the DC power supplies (DCs) for compensating the DC drifts \cite{Mitsugi96,Nagata04,Nagata04b,Salvestrini11} of LN MODs.
We used a commercially available polarimeter (PM1) to monitor the polarisation states at the maximum sampling rate of 400 Hz, and then, DC voltages were applied to MOD1 and MOD2 to compensate the DC drifts by feedback control through a computer.
The sampling rate is slow enough to allow the bias-tees to respond through the DC port, such that we could stabilise the Poincar\'e rotator against the DC drifts, while RF signals were transmitted after the amplifications through the bias-tees.
This combination of the rotator and the phase-shifter finalises the set-up of the Poincar\'e rotator.

The final set-up is to prepare the measurement system, since a conventional polarimeter was not fast enough to observe the modulation signals out of the Poincar\'e rotator.
We have followed a standard set-up of a retarder-polariser configuration to analyse the polarisation state \cite{Gil16,Goldstein11}.
The retarder is made of a waveplate, whose SA is aligned horizontally, whose operator is given by 
\begin{eqnarray}
\hat{\mathcal{R}}_1^{\prime} (-\delta)
=
\left(
\begin{array}{cc}
1 & 0 \\
0 & {\rm e}^{-i  \delta}
  \end{array}
\right), 
\end{eqnarray}
where the phase delay of $\delta$ is $\pi$, $\pi/2$ and $0$ for a HWP, a QWP, and no plate, respectively.
Then, we will use a rotated polariser, which is physically rotated with the amount of ${\it \Delta \Psi}$ from the horizontal direction for the anti-clock-wise direction, and the operator becomes 
\begin{eqnarray}
\hat{\mathcal{P}} (\Delta \phi)
=
\frac{1}{2}
\left(
\begin{array}{cc}
1+\cos(\Delta \phi) & \sin(\Delta \phi) \\
\sin(\Delta \phi) & 1-\cos(\Delta \phi)
  \end{array}
\right), 
\end{eqnarray}
where $\Delta \phi = 2 {\it \Delta \Psi}$ is amount of the rotation on the Poincar\'e sphere.
These operators were applied to a general input state for polarisation in complex electric fields \cite{Max99,Jackson99,Yariv97,Parker05,Chuang09,Grynberg10,Gil16,Goldstein11,Hecht17,Pedrotti07,
Saito20a,Saito22g}, 
\begin{eqnarray}
|{\rm Input} \rangle
=
\left(
\begin{array}{c}
E_x \\
E_y
\end{array}
\right), 
\end{eqnarray}
and the output state becomes
\begin{eqnarray}
|{\rm Output} \rangle
&=&
\hat{\mathcal{P}} (\Delta \phi)
\hat{\mathcal{R}}_1^{\prime} (-\delta)
|{\rm Input} \rangle \\
&=&
\left(
\begin{array}{c}
(1+\cos(\Delta \phi))E_x +  {\rm e}^{-i  \delta} \sin(\Delta \phi)  E_y  \\
\sin(\Delta \phi)  E_x + (1-\cos(\Delta \phi)) {\rm e}^{-i  \delta} E_y 
\end{array}
\right) \nonumber \\
\end{eqnarray}
The output intensity, $I(\delta,\Delta \phi)$, is obtained \cite{Max99,Jackson99,Yariv97,Gil16,Goldstein11} as
\begin{eqnarray}
I(\delta,\Delta \phi)
=&&
\langle {\rm Output}
| {\rm Output} \rangle \\
=&&
\frac{1}{2}S_0
+
\frac{1}{2} \cos(\Delta \phi) S_1
+
\frac{1}{2} \sin(\Delta \phi) \cos \delta S_2
\nonumber \\
&&
+
\frac{1}{2} \sin(\Delta \phi) \sin \delta S_3
,
\end{eqnarray}
where the Stokes parameters are given by
\begin{eqnarray}
S_0 &=& |E_x|^2+|E_y|^2=I_0+I_1 \\
S_1 &=& |E_x|^2-|E_y|^2=I_0-I_1 \\
S_2 &=& E_x E_y^{*}+E_x^{*} E_y=2I_2-I_0-I_1 \\
S_3 &=& i(E_x E_y^{*}-E_x^{*} E_y)=2I_3-I_0-I_1
,
\end{eqnarray}
such that we just need to monitor intensities of 4 configurations as
\begin{eqnarray}
&&I(0,0)=\frac{1}{2}(S_0+S_1)\equiv I_0 \\
&&I(0,\pi)=\frac{1}{2}(S_0-S_1)\equiv I_1 \\
&&I(0,\pi/2)=\frac{1}{2}(S_0+S_2)\equiv I_2 \\
&&I(\pi/2,\pi/2)=\frac{1}{2}(S_0+S_3)\equiv I_3 
.
\end{eqnarray}
$I_0$, $I_2$, and $I_3$ monitor $|E_x|^2$, $|E_{\rm D}|^2$, and $|E_{\rm L}|^2$, respectively, where $E_{\rm D}$ and $E_{\rm L}$ are complex electric fields for D and L states, respectively, since $I_2=(|E_{\rm D}|^2-|E_{\rm A}|^2)/2$ and $I_3=(|E_{\rm L}|^2-|E_{\rm R}|^2)/2$, where $E_{\rm A}$ and $E_{\rm R}$ are complex electric fields for A and R states, respectively.
From this classical formula \cite{Max99,Jackson99,Yariv97,Gil16,Goldstein11}, we understand that we need intensities of H, V, D, and L states, after splitting the ray into 4 rays without changing the polarisation states.
We used a filter to block the A state, simply by rotating the linear polariser at $45^{\circ}$ to pass only for the D-state.
If we use a QWP, whose SA is aligned horizontally, we can also set-up an R-filter, which pass only for the L-state.

The experimental set-up of the bespoke polarimeter is shown in Fig. 3 (the cyan part at the bottom).
Here, we have used Er-doped fibre amplifier (EDFA) to amplify the signal before analysing the polarisation state.
We used a polarisation independent EDFA to amplify the input with an arbitrary polarisation state without changing it at the output.
This is not violating with a non-cloning theorem \cite{Wootters82,Dieks82}, which prohibits the copying of a quantum state by linear operations.
In fact, stimulated emission processes must take place under the population inversions \cite{Yariv97,Parker05,Chuang09,Grynberg10}, which correspond to cloning of photons with phase coherence to maintain the polarisation state.
Under an EDFA, extra energies were provided to coherent photons by optical pumping, such that the linear operation condition for validating a non-cloning theorem were not met.
Cloning of photons under an EDFA or a laser was well-established, and otherwise, lasing of photons would not be achieved.
An EDFA simply generates photons, which reside in the same quantum states with the input photons through stimulated emissions, and there is no conceptual difficulty nor a challenge to a law in physics to copy local quantum states during stimulated emission processes.
We have used the EDFA under the pump current of 750mA, whose induced a gain of 28 dB.
The intensities of H, V, D, and L states ($I_{\rm H}$, $I_{\rm V}$, $I_{\rm D}$, and $I_{\rm L}$, respectively) were measured by 4 photo-diodes (PDs), whose bandwidth was 20GHz, and output signals were monitored by a oscilloscope (OSC), whose bandwidth was 500MHz.

\subsection{Phase-Shifter}
First, we have examined the operation of a phase-shifter without operating a rotator at 10 MHz for the duty cycle of 10 \%.
We used simple sine waves for the inputs (Supplementary Fig. 1) to control the phase-shifter, and examples of output signals for the phase-shifter is shown in Supplementary Fig. 2.
We obtained the Stokes parameters from the output signals, which were obtained by applying the half of the voltage required for the $\pi$ phase-shift ($V_{\pi}=0.56$ V) to the LN modulator, as shown in Fig. 4.
Our duty cycle was 10 \%, such that the polarisation state was mostly found at the input state of the D state (Fig. 4 (b)), while the minimum value of $S_2$ was located around 0 during sine wave modulations.
On the other hand, the applied quarter voltage of $V_{\pi}/2=0.28$ V, was enough to bring the D state to L and R state along the parallels, such that the minimum and maximum values of $S_3$ was 1 and -1, respectively (Fig. 4 (c)).
The $S_1$ values should be constant, ideally, but they were fluctuating around 0.
We estimated the maximum deviations of the order of $\pm 10^{\circ}$ could be found in our set-up \cite{Saito22g}, such that the observed fluctuations were satisfactory.

\begin{figure}[h]
\begin{center}
\includegraphics[width=8cm]{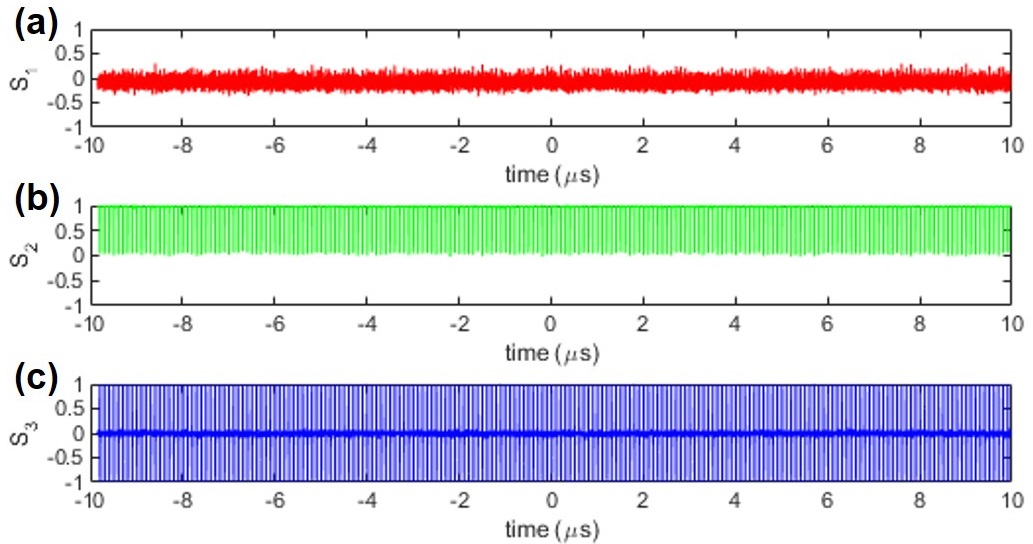}
\caption{
Stokes parameters (a) $S_1$, (b) $S_2$, and (c) $S_3$, controlled by the phase-shifter, at the maximum input voltage of $V_{\pi}/2=2.8$ V applied to the modulator.
The input ray was in the D state, which was located at $(S_1,S_2,S_3)=(0,1,0)$, and the polarisation state was modulated at 10 MHz with a duty of 10 \%.
From the scanned time scale from -10 ${\rm \mu m}$ to 10 ${\rm \mu m}$, 200 sine pulses were applied. 
}
\end{center}
\end{figure}

If we focus on the modulation region, we can clearly confirm the expected changes in the hemisphere along the parallels in the HV basis, the trajectory on the Poincar\'e sphere is located within the $S_2$-$S_3$ plane at $S_1=0$  (the yellow curve of Fig. 5 (b)).
We confirmed that the output signals of $S_3$ (the blue curve of Fig. 5 (a)) reflected the shape of the input signal of the sine form, including the shape of the ripples (Supplementary Fig. 1).
Trajectories of $S_2$ had minimum twice during the 1 cycle, since $S_2$ vanishes at both L and R states, located north and south poles for a LR basis.

\begin{figure}[h]
\begin{center}
\includegraphics[width=8cm]{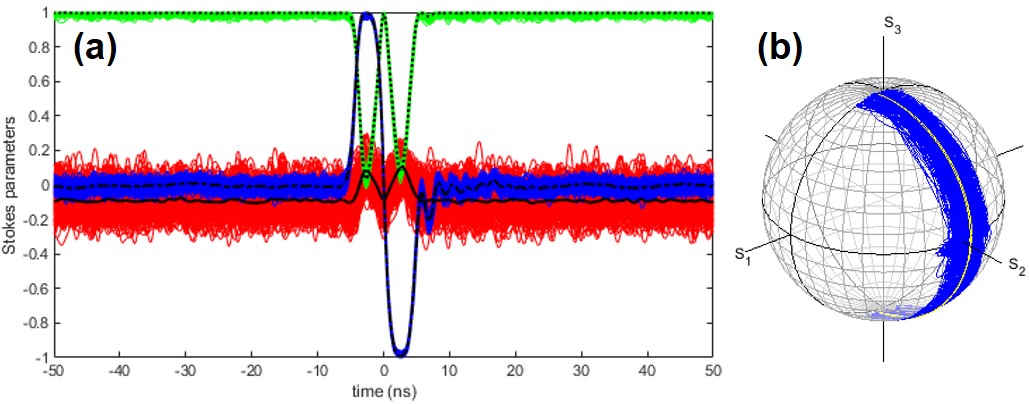}
\caption{
Phase-shifter operation at the maximum input voltage of $V_{\pi}/2=2.8$ V applied to the modulator.
(a) Stokes parameters of $S_1$ (red), $S_2$ (green), and $S_3$ (blue). 
The averaged curves are also shown for $S_1$ (sold line), $S_2$ (dashed line), and $S_3$ (dotted line), respectively.
(b) Poincar\'e sphere. 
The original data for 200 pulses were plotted in blue, while the averaged trajectory was plotted in yellow.
}
\end{center}
\end{figure}

\begin{figure}[h]
\begin{center}
\includegraphics[width=8cm]{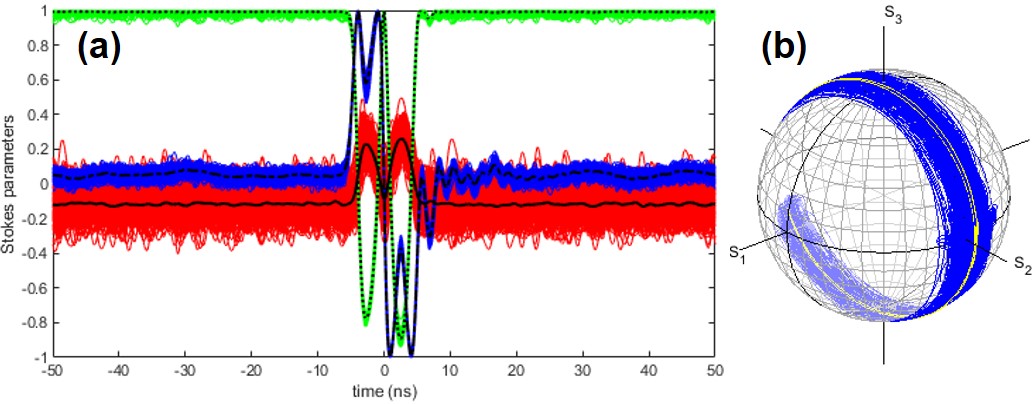}
\caption{
Phase-shifter operation at the maximum input voltage of 5.1 V, which is just below $V_{\pi}=5.6$ V, required for 1 circulation.
(a) Stokes parameters of $S_1$ (red), $S_2$ (green), and $S_3$ (blue). 
(b) Poincar\'e sphere. 
}
\end{center}
\end{figure}

We have also examined the phase-shifter by generating 0.4 V from the AWG, which corresponds to apply the maximum voltage of $0.4$ V $\times 12.7 =5.1$ V to a modulator.
This voltage is just below $V_{\pi}=5.6$ V for 1 circulation on the Poincar\'e sphere, which is exactly what we observed (Fig. 6 (c)).
In Stokes parameters, $S_3$ values have dips to show over-rotations beyond L and R states from the input of the D state (the blue curve of Fig. 6 (a)).
Conversely, we confirmed that the dips to show the over-rotation disappeared at the AWG voltage of 0.22 V by changing the voltage at the step of 0.01 V, and the extracted voltage of $V_{\pi}=5.6$ V was consistent with the specification of our LN phase modulator.
The 2 dips in $S_2$ values were approached to the minimum of -1, but the voltage was not enough to arrive the A state, as expected (the green curve of Fig. 6 (a)).

\subsection{Rotator}
We have also examined the rotator performance at 10 MHz without using the phase-shifter (Supplementary Fig. 3, Figs. 7-9).
We confirmed similar performance with the phase-shifter, while the rotation axis was successfully rotated from $S_1$ to $S_3$, which is evidenced by the constant $S_3$ value in Fig. 7 (c).
In Fig. 7, $S_1$ was fully changed, showing the changes from H to V states, while $S_2$ changed between 1 and 0, showing the hemisphere scanning, consistent with the quarter voltage of $V_{\pi}/2$ to the modulator.
In fact, the rotator changed the polarisation state within the $S_1$-$S_2$ hemisphere of $S_2>0$ at $S_3=0$, as shown in Fig. 8.
At higher voltages (Fig. 9), we also confirmed the expected over-rotations beyond H and V states (the red curves in Fig. 9 (a)).

\begin{figure}[h]
\begin{center}
\includegraphics[width=8cm]{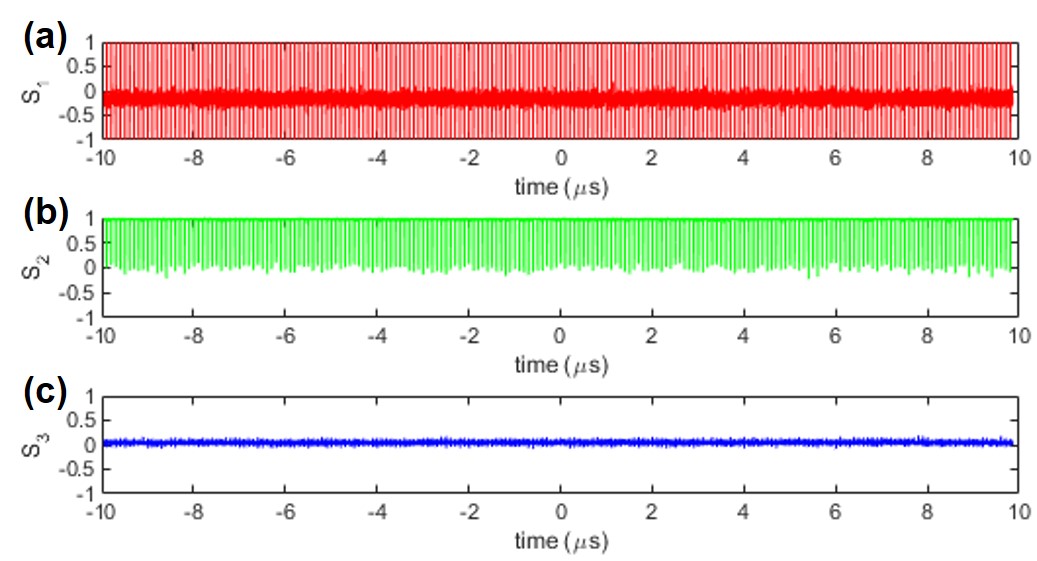}
\caption{
Stokes parameters (a) $S_1$, (b) $S_2$, and (c) $S_3$, controlled by the rotator, at the maximum input voltage of $V_{\pi}/2=2.8$ V applied to the modulator.
}
\end{center}
\end{figure}

\begin{figure}[h]
\begin{center}
\includegraphics[width=8cm]{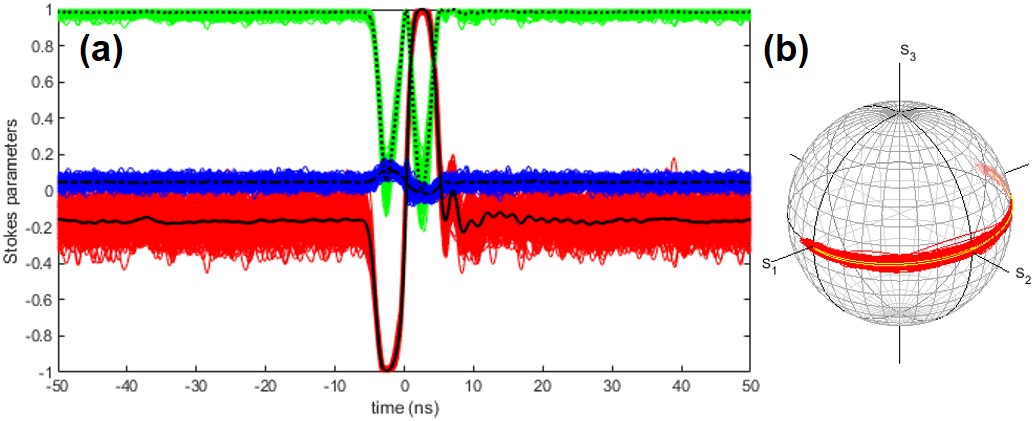}
\caption{
Rotator operation at the maximum input voltage of $V_{\pi}/2=2.8$ V applied to the modulator.
(a) Stokes parameters of $S_1$ (red), $S_2$ (green), and $S_3$ (blue). 
The averaged curves are also shown for $S_1$ (sold line), $S_2$ (dashed line), and $S_3$ (dotted line), respectively.
(b) Poincar\'e sphere. 
The original data for 200 pulses were plotted in red, while the averaged trajectory was plotted in yellow.
}
\end{center}
\end{figure}

\begin{figure}[h]
\begin{center}
\includegraphics[width=8cm]{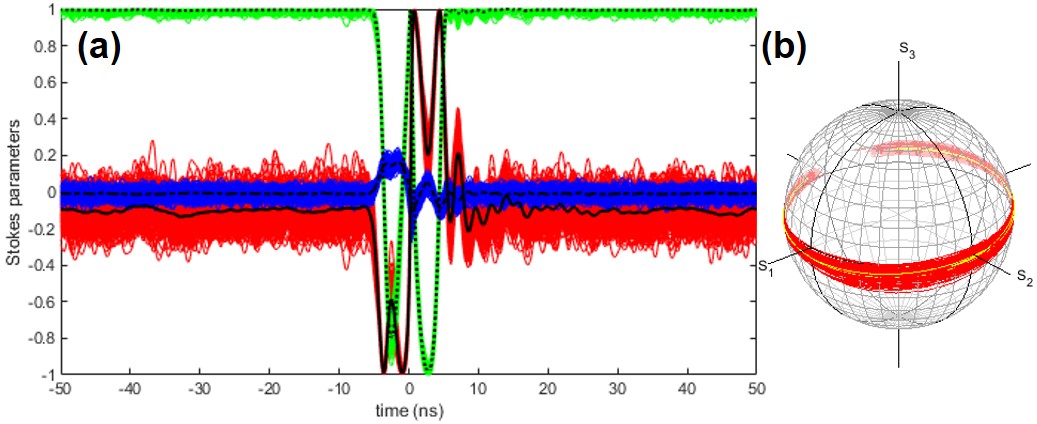}
\caption{
Rotator operation at the maximum input voltage of 5.1 V.
(a) Stokes parameters of $S_1$ (red), $S_2$ (green), and $S_3$ (blue). 
(b) Poincar\'e sphere. 
}
\end{center}
\end{figure}

These experiments on the phase-shifter and the rotator confirmed that we could construct $U(1)$ operators along both meridians and parallels, respectively, on the Poincar\'e sphere, as theoretically expected in Fig. 1.
$U(1)$ operations for 2 orthogonal rotational axes are prerequisite for the construction of proper $SU(2)$ operations to allow arbitrary rotations on the sphere.

\section{Results}
Next, we have examined the sequential operations of Poincar\'e rotator by combining the rotator and the phase-shifter to show the arbitrary $SU(2)$ operations to single bits.
In order to demonstrate the capabilities of the Poincar\'e sphere, we have tried to access the multiple points on the sphere by changing the polarisation states.
If we can resolve each points on the Poincar\'e sphere, we can use them for transmitting information to distinguish from different polarisation states, as an application. 
More fundamentally, operations of our Poincar\'e rotator are exactly the same as those expected for $SU(2)$ operations in quantum mechanics, while our quantum bits were based on macroscopic number of photons from a standard laser source.
We thought it was not trivial to realise expected rotations, which were successfully confirmed for single-qubits rotations.

For calibrating the synchronised operations, we found that the pulse delay of 27 ns for the signal modulated by the rotator to arrive at the modulator for the phase-shifter, which was consistent with the corresponding optical path length of 5.6 m in fibres for our experiments.
The delay was introduced in the input waveforms for AWGs to allow the sequential rotations.
In previous experiments, shown above, we confirmed the averaging over 200 scans improved the signal-to-noise ratio significantly (yellow curves in Figs. 5, 6, 8, and 9), 
Therefore, we have decided to take an analogue average over 128 curves, during the data acquisitions by synchronising the signals in the oscilloscope through triggers from AWGs for the data, shown below.

In order to rotate the polarisation states on the full Poincar\'e sphere, we need to scan the polar angle of $\gamma$ from $-\pi/2$ to $\pi/2$ and the azimuthal angle of $\delta$ from $-\pi$ to $\pi$, since the input polarisation is the D state, located at $(S_1,S_2,S_3)=(0,1,0)$.
We usde the rotator, first, to scan along meridians, followed by the phase-shifter to scan along parallels (Fig. 3).
The input voltage of $V_1$ was generated from AWG1 and applied to MOD1, and the input voltage of $V_2$ was generated from AWG2 and applied to MOD2 after the voltage amplifications (Fig.3).
The maximum amplitude of $|V_1|$ was set to be 0.22 V to generate $\pm V_{\pi}/2$ as the output voltage after the amplifier, while the maximum amplitude of $|V_2|$ was set to be 0.44 V to generate $\pm V_{\pi}$ as the output voltage after the amplifier.

We have examined for realising $4 \times 4=16$, $8 \times 8=64$, and $10 \times 10=100$ polarisation states by operating the Poincar\'e rotator at 10 MHz with the duty cycle of 10 \%, which corresponds to the pulse duration of 10 ns for the clock of 100 ns in periods.
We have also demonstrated to show the structures of Fullerene C$_{60}$ and the earth on the Poincar\'e sphere at 5 MHz with the duty cycle of 5 \%, which corresponds to the pulse duration of 10 ns for the clock of 200 ns in periods.

\subsection{$4 \times 4=16$ bits operation}

\begin{figure}[h]
\begin{center}
\includegraphics[width=8cm]{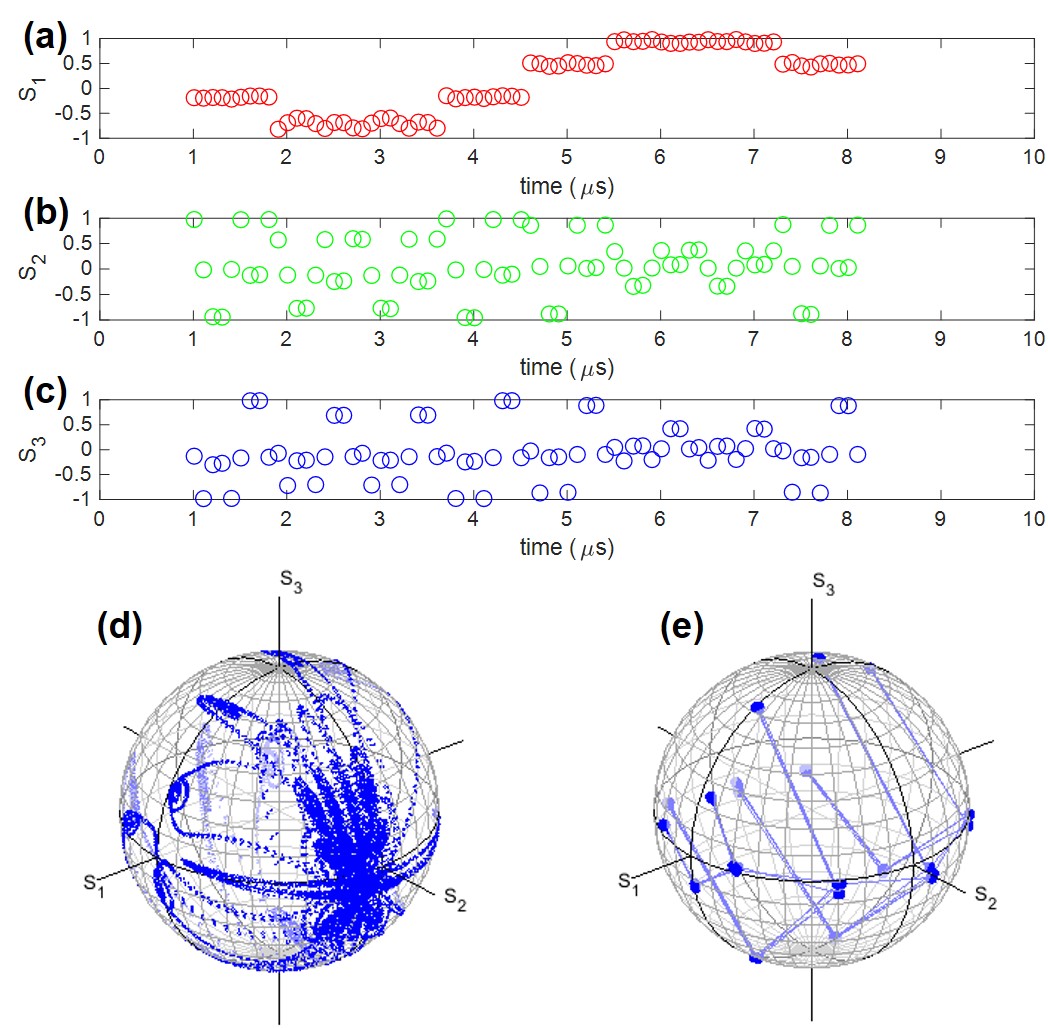}
\caption{
$4 \times 4=16$ bits operation by the Poincar\'e rotator.
Stokes parameters of (a) $S_1$, (b) $S_2$, and (c) $S_3$ at central points in the pulse duration of 10 ns with the clock of 100 ns.
(d) Full trajectories of polarisation states on the Poincar\'e sphere.
(e) Polarisation states at the central time in the pulse. Lines are just guides to the eye by connecting the nearest neighbour points of pulses.
}
\end{center}
\end{figure}

The input waveforms for $4 \times 4=16$ polarisation states and original output signals are shown in Supplementary Fig. 4.
This means that we have prepared 4 different voltages of $V_1$ for $\gamma$ and 4 different voltages of $V_2$ for $\delta$ (Supplementary Figs. 4 (a) and 4 (b)).
The full trajectories of the Stokes parameters are shown in Fig. 10 (d).
The input polarisation state was the D state and we used return-to-zero operations for modulators, such that the polarisation state is mostly found in the D state throughout the scan.
We can also recognise that the polarisation states were changed continuously from the D state to the targetted points on the Poincar\'e sphere.

We have taken the central data points of observed Stokes parameters within the pulse duration of 10 ns for each pulse, in conjunction with the clock of 100 ns (Figs. 10 (a)-(c)).
We can clearly recognise that 4 different levels of $S_1$ values were successfully realised by the rotator (Fig. 10 (a)), which were not disturbed by the following phase-shifter operations.
The successive 4 different modulations by the phase-shifter were less clear in the values of $S_2$ and $S_3$, since the phase-shifter induced full rotations along parallels.
However, if we plot the extracted central points on the Poincar\'e sphere, we can clearly confirm the expected modulations (Fig. 10 (e)).
We see that the phase-shifter actually realised 4 points in parallel to the $S_2$-$S_3$ plane without changing the values of $S_1$, and we confirmed that 16 points on the Poincar\'e sphere were distinguishable.

\subsection{$8 \times 8=64$ bits operation}

\begin{figure}[h]
\begin{center}
\includegraphics[width=8cm]{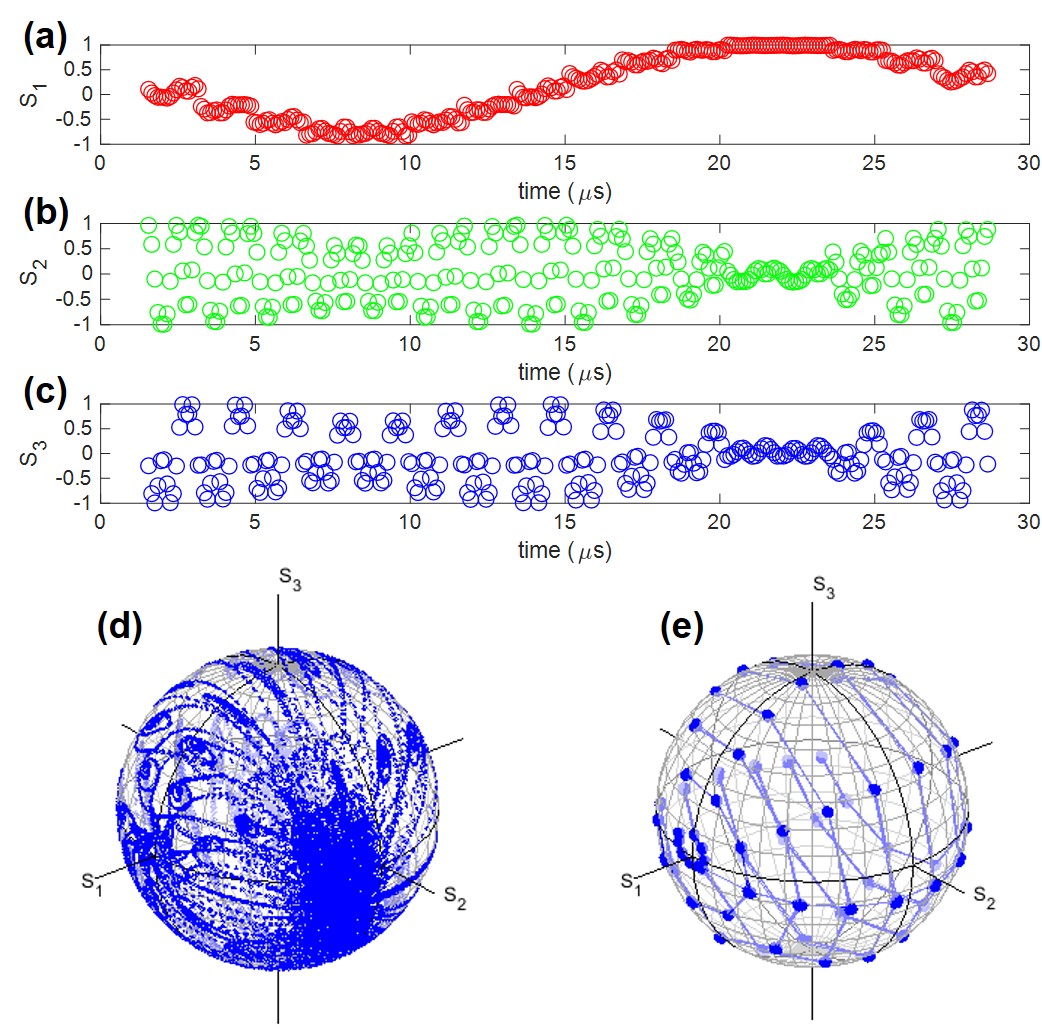}
\caption{
$8 \times 8=64$ bits operation by the Poincar\'e rotator.
Stokes parameters of (a) $S_1$, (b) $S_2$, and (c) $S_3$ at central points in the pulse duration of 10 ns with the clock of 100 ns.
(d) Full trajectories of polarisation states on the Poincar\'e sphere.
(e) Polarisation states at the central time in the pulse. 
}
\end{center}
\end{figure}

Next, we have examined to realise $8 \times 8=64$ polarisation states on the Poincar\'e sphere.
The input waveforms and original Stokes parameters are shown in Supplementary Fig. 5, and extracted points at the centre of the pulses are shown in Fig. 11.
Here, we have gradually modulated from the south hemisphere ($S_1<0$) to the north hemisphere ($S_1>0$) for meridians in a HV basis by the rotator, as seen from Fig. 11 (a), while the parallels were modulated by the phase-shifter for each $S_1$ level, as seen in Fig. 11 (e).
In fact, the values of the radius during the phase-shifter rotations became smaller as the $S_1$ values approached to north ($S_1=1$) and south ($S_1=-1$) poles, which were expected theoretically in Fig. 2 (b). 
We recognised 8 levels in the 8 circles, such that total 64 points were distinguishable.
On the full trajectories (Fig. 11 (d)), it is very hard to distinguish targetted points on the Poincar\'e sphere, and it is important to synchronise with clocks, as is always true for any signal transmissions.

\subsection{$10 \times 10=100$ bits operation}

\begin{figure}[h]
\begin{center}
\includegraphics[width=8cm]{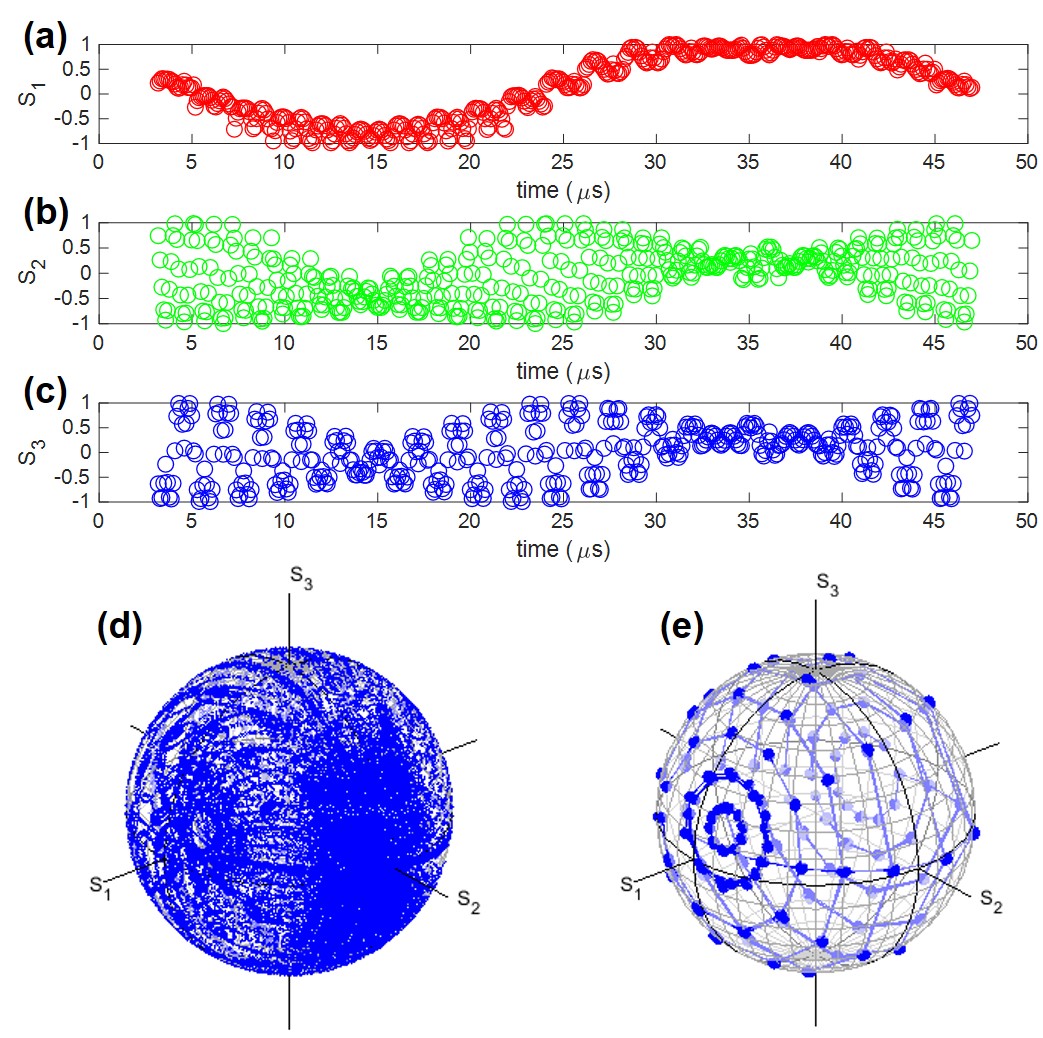}
\caption{
$10 \times 10=100$ bits operation by the Poincar\'e rotator.
Stokes parameters of (a) $S_1$, (b) $S_2$, and (c) $S_3$ at central points in the pulse duration of 10 ns with the clock of 100 ns.
(d) Full trajectories of polarisation states on the Poincar\'e sphere.
(e) Polarisation states at the central time in the pulse. 
}
\end{center}
\end{figure}

Finally for the multi-bits realisation, we have also tried for realising $10 \times 10=100$ polarisation states on the Poincar\'e sphere, as shown in Supplementary Fig. 6 and Fig. 12. 
The full trajectories almost completely filled the entire Poincar\'e sphere, while multi-bits are recognised at specific timing on the clock.
We could barely recognise 100 points, however, at the present experimental status on the expected deviations of $\pm 10$ \% in angles, this experiment was almost maximum on multiplicities, we could expect.  
In order to accommodate more bits, we need to improve the fidelities by optimising optical components and improving the alignment accuracies.
Moreover, the radius during the rotation by the phase-shifter became even smaller (Fig. 12 (e)), such that it was difficult to distinguish different polarisation points near north and south poles, due to the unequal spacing between polarisation points.

\subsection{Fullerene C$_{60}$ on the Poincar\'e sphere}

\begin{figure}[h]
\begin{center}
\includegraphics[width=8cm]{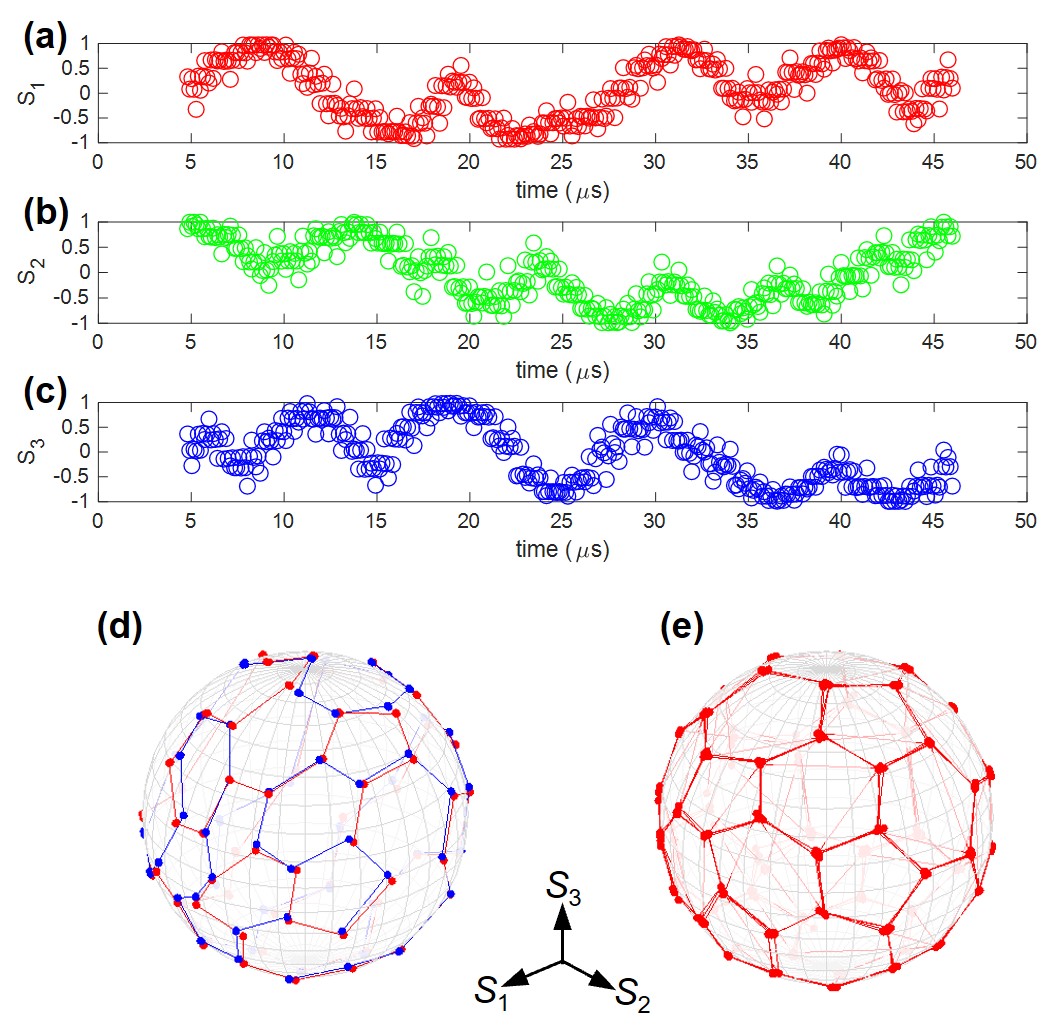}
\caption{
Fullerene C$_{60}$ on the Poincar\'e sphere as polarisation states of pulses.
Stokes parameters of (a) $S_1$, (b) $S_2$, and (c) $S_3$, extracted at the centre of the pulse.
(d) C$_{60}$ realised by left (red) and right (blue) Hamilton paths, respectively.
(e) C$_{60}$ realised by a left Hamilton path together with 3 nearest neighbour visits of the vertexes. 
For (d) and (e), experimental polarisation states are shown on the Poincar\'e sphere as points to represent vertexes of C$_{60}$. 
The nearest points during the scan were connected by lines as bonds to represent nearest-neighbour visits as guides to the eye.
}
\end{center}
\end{figure}

Having realised about 100 different polarisation states on the Poincar\'e sphere, we considered how to improve the visibility to distinguish points on the sphere, which would be important for transmitting polarisation-multiplexed information or any other applications to distinguish polarisations.
We should keep sufficient distances among neighbouring points on the sphere, and each point should be symmetric and equivalent.
For that purpose, we think the structure of Buckminsterfullerene (fullerene, C$_{60}$) \cite{Kroto85} is ideal, since the fullerene C$_{60}$ has a spherical symmetry, and all points are  equivalent to each other as a vertex for 2 pentagon and 1 hexagon.
The coordinates of the vertexes for C$_{60}$ were calculated by a graph theory \cite{Pemmaraju03,Attaway18}.
In order to draw a C$_{60}$ molecule on the Poincar\'e sphere for a sequential manner, connecting neighbouring bonds, we have calculated a Hamilton path, which is a path to connect all vertexes just by visiting each vertex once \cite{Pemmaraju03,Attaway18}, for C$_{60}$.
Indeed, there exists several Hamilton paths for C$_{60}$, and we used a Hamilton path for left circulation as well as the mirror-symmetric right circulation.
We have also tried to connect all bonds during the scanning, since both left and right Hamilton paths do not cover all the bonds.

The input waveforms for $V_1$ and $V_2$ and observed Stokes parameters for the left Hamilton path of C$_{60}$ are shown in Supplementary Fig. 7.
We have operated at a reduced speed of 5 MHz with a duty of 5 \% for the pulse duration of 10 ns, to clarify the vertexes of C$_{60}$ without interferences between pulses, 
From the trajectories of Stokes parameters (Supplementary Figs. 7 (c)-(e)), we have extracted central points at each pulse and plotted on the Poincar\'e sphere (the red points of Fig. 13 (d)).
We confirmed the same topological structure with C$_{60}$, shown on the Poincar\'e sphere.
We have also tried to make a mirror-imaged path of the right Hamilton path (the blue points of Fig. 13 (d)), confirming that we could reproduce the same vertexes in a different path.
We see that there are several bonds, missing in these paths, due to the restrictions to visit each vertex only once.

In order to confirm that we could access to each neighbouring vertexes, we have also prepared another path to connect all bonds (Supplementary Fig. 8).
Here, we used a left Hamilton path to access the vertex, and at each vertex we have added 3 nearest neighbouring vertexes to visit before going to the next.
In this case, $60 \times 3 =180$ pulses were employed for each scan.
The observed Stokes parameters at the centre of the pulses were shown in Figs. 13 (a)-(c) and plotted on the Poincar\'e sphere, as shown in Fig. 13 (f).
We can successfully confirm that each experimental point on the Poincar\'e sphere represented the vertex of C$_{60}$ as a vertex of 2 pentagons and 1 hexagon. 
Therefore, all points are topologically equivalent to each other, respecting the spherical symmetry.
Therefore, the pulse sequence for using C$_{60}$ structures would be more preferable than the multi-bit sequence for $8\times8=64$ bits, presented in the former sub-section B.

In the actual experiments, we admit to recognise, of course, the bondings were not perfect, in the sense that some distances were different, such that photonic C$_{60}$ on the Poincar\'e sphere was distorted. 
This was particularly true for the states opposite to the input state of the D state, since larger voltages were required for realising the points near the A state for the phase-shifter.
For equalising the points, it will be possible for optimising the input signals by taking care of the frequency responses of our devices, which will be important for higher speed operations.
For applications in communications, it is possible to combine our polarisation controls, together with the amplitude controls like QAM and PAM.
If we use the multiplexing using the structure of C$_{60}$, the data transmission rate will be potentially increased as high as 60 times than the current format.

\subsection{Coastlines of the earth on the Poincar\'e sphere}

\begin{figure}[h]
\begin{center}
\includegraphics[width=8cm]{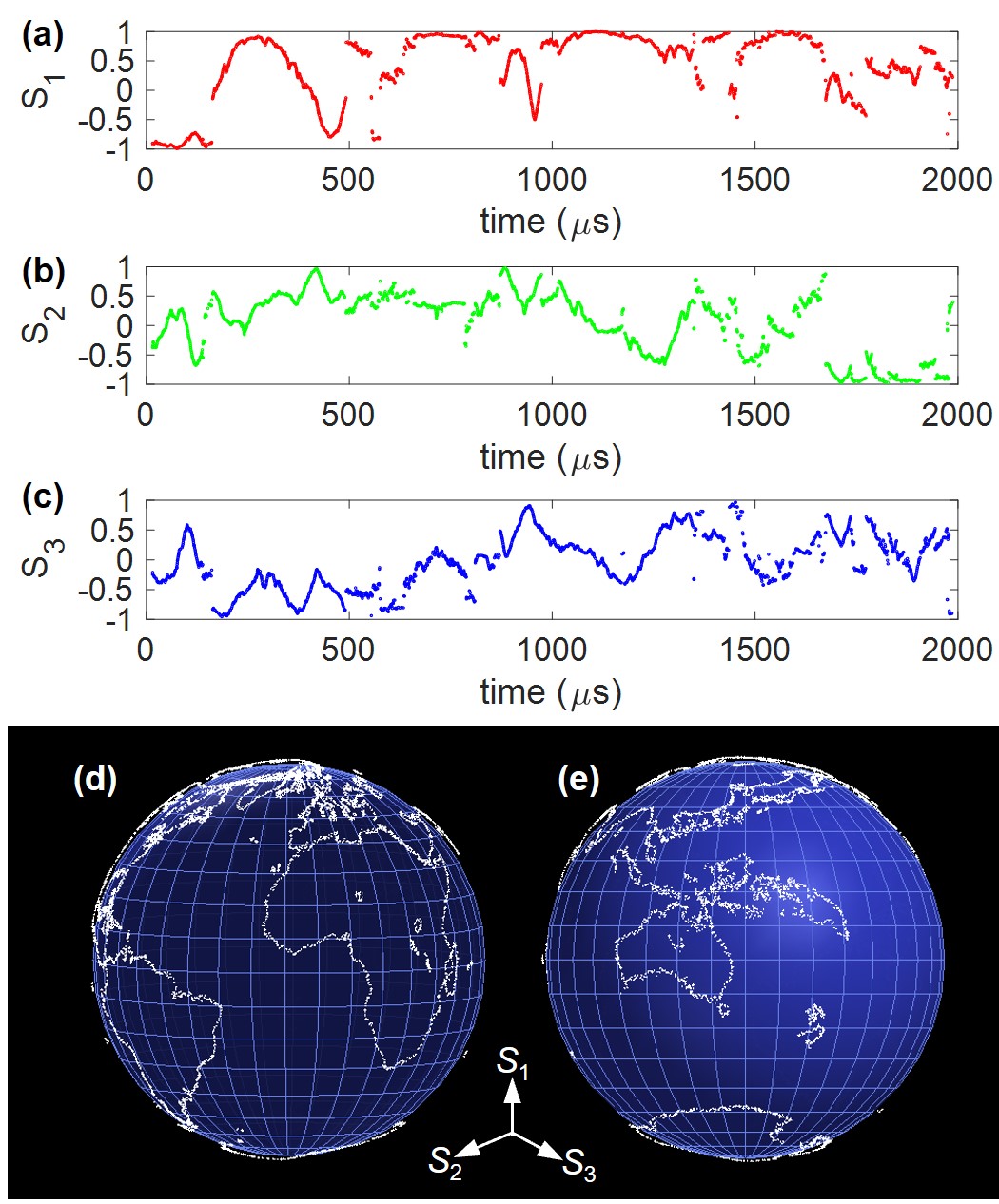}
\caption{
Coastlines of the earth on the Poincar\'e sphere as polarisation states of pulses.
Stokes parameters of (a) $S_1$, (b) $S_2$, and (c) $S_3$, extracted at the centre of the pulse.
(d) Asia and the Pacific region.
(e) Europe, Africa, and the Atlantic region. 
Please note polarisation states at $S_1=\pm 1$ represent north and south poles in a HV basis, respectively.
}
\end{center}
\end{figure}

As a final demonstration, we have tried to represent the coastlines of the earth by the polarisation states on the Poincar\'e sphere (Fig. 14).
We have used coastline data in a package of Matlab \cite{Attaway18}, which is made of 9,865 points and we converted the coordinates to the data for the Poincar\'e rotator (Supplementary Figs. 9 (a) and (b)).
We have operated at 5 MHz with the duty cycle of 5 \% for the pulse duration of 10 ns at each point.
Trajectories on Stokes parameters are shown in Supplementary Figs. 9 (c)-(d), and the central points at each pulse with the clock of 200 ns are shown in Figs. 14 (a)-(c), and they were plotted on the Poincar\'e sphere (Figs. 14 (d) and (e)).
We can recognise a familiar view of the earth, which looked better than we expected, considering the estimated deviations of $\pm 10$ \%.
This was supported by our feedback system to stabilise the DC drift (Fig. 3).
We admit to see some distortions of coastlines, which were purely originated from our poor performance of the Poincar\'e rotator at this stage and we sincerely apologise the potential concern on the drawing.
This demonstration at least shows a capability to represent points on a sphere by using the polarisation states on the Poincar\'e sphere.
There are many other candidates, which could be suitable to represent on a sphere.
The proposed Poincar\'e rotator is a device to represent them as polarisation states of coherent photons, such that it will be useful to transmit information or to store information in a fibre as flying bits.
In fact, a polarisation state described by a $SU(2)$ state can accommodate huge information for the future. 
At the current status, it is limited up to 60 to 100 due to our experimental set-up with a fidelity of $\sim 90$ \%.
If we could improve the fidelity to 99.9 \%, we can store 1k different states in a $U(1)$ group for the phase and 1M different states in an $SU(2)$ group for the phase and the amplitude.
For example, we can pick up 1 city out of 1 M cities on the earth,  using the polarisation state of a $SU(2)$ group, which is completely beyond conventional binary digital data processing.
Therefore, it is promising to explore polarisation degrees of freedom more for information and communication technologies, inspired by the quantum aspects of the polarisation states.

\section{Discussions}

Our original motivation of this study was to explore macroscopic quantum nature of polarisation states of coherent photons, described by an $SU(2)$ theory \cite{Saito20a,Saito20b,Saito20c,Saito20d,Saito20e,Saito21f,Saito22g}.
As shown above, polarisation states of coherent states could be described by a simple quantum mechanics of a 2 level system \cite{Jones41,Fano54,Baym69,Sakurai67,Sakurai14,Max99,Jackson99,Yariv97,Gil16,Goldstein11}, regardless of a macroscopic number of photons in each pulse.
Our treatment of polarisation states are purely based on a quantum-mechanical $SU(2)$ theory, due to the macroscopic coherency of photons, exhibiting macroscopic quantum phenomena of the type-I similar to BEC \cite{Takagi02}.
For coherent photons, we could utilise these $SU(2)$ degrees of freedom, due to the absence of charge for a photon, and we could even construct quantum-mechanical $SU(2)$ operators, which we named as the Poincar\'e rotator, which allowed arbitrary rotations on the Poincar\'e sphere.
Therefore, we think it is appropriate to represent a qubit by a polarisation state of a laser pulse, such that we call it as a macroscopic qubit with a $SU(2)$ character.
Here, we must emphasise that we could realise only a single-qubit operation only, and a two-qubit operation is far beyond the scope of this paper.
The macroscopic two-qubit operation might be prohibited by some principles, and we cannot discuss any further on this potential.

Instead, we would like to discuss on the measurements of the polarisation states, since  observations are important in understanding quantum-mechanics \cite{Dirac30,Baym69,Sakurai14,Max99,Jackson99,Yariv97,Gil16,Goldstein11,
Wootters82,Dieks82}.
We have previously showed that the Stokes parameters are expectation values of spin angular momentum for coherent photons \cite{Saito20a,Saito20c}, and we could experimentally set-up to measure these spin components.
As described above, we have used EDFA to amplify the signals before the measurements, and we could observe expected operations by the Poincar\'e rotator, such that cloning of polarisation states with $SU(2)$ degrees of freedom seems to be successfully realised in experiments without any issues.
We need to understand this does not violate with the no-cloning theorem, which prevents to make a copy of an original state by a unitary operation \cite{Wootters82,Dieks82}.

In order to confirm the reason why we could amplify the polarisation states without destroying them, it is useful to consider coherent states, \cite{Grynberg10,Fox06,Parker05,Saito20a}, $|\alpha_{\rm H},\alpha_{\rm V}\rangle =|\alpha_{\rm H}\rangle | \alpha_{\rm V}\rangle$, which is given by 
\begin{eqnarray}
|\alpha_{\rm H} \rangle
&=&{\rm e}^{-\frac{|\alpha_{\rm H}|^2}{2}}
{\rm e}^{\alpha_{\rm H} \hat{a}_{\rm H}^{\dagger}}
|0\rangle \\
|\alpha_{\rm V} \rangle
&=&{\rm e}^{-\frac{|\alpha_{\rm V}|^2}{2}}
{\rm e}^{\alpha_{\rm V} \hat{a}_{\rm V}^{\dagger}}
|0\rangle, 
\end{eqnarray}
where $ \hat{a}_{\rm H}^{\dagger}$ and $ \hat{a}_{\rm V}^{\dagger}$ are creation operators for H and V states, respectively, to satisfy the bose commutation relationships.
The parameters are obtained as $\alpha_{\rm H} =\sqrt{N_{\rm H}}=\sqrt{N} \cos \alpha$, $\alpha_{\rm V}=\sqrt{N_{\rm V}} {\rm e}^{i \delta} =\sqrt{N} \sin \alpha {\rm e}^{i \delta}$, where we consider photons in each pulse and the average numbers of photons for each polarisation are $N_{\rm H}$ and  $N_{\rm V}$, and the total number of photons in the pulse is  $N=N_{\rm H}+N_{\rm V}$.
The spin angular momentum operator, $\hat{\bf S}$, is given by a spinor representation of the field operator $\bm{\hat{\psi}}^{\dagger}=(\hat{a}_{\rm H}^{\dagger},\hat{a}_{\rm V}^{\dagger})$ and $\bm{\hat{\psi}}$ as 
\begin{eqnarray}
\hat{\bf S}
&=&
\hbar  
\bm{\hat{\psi}}^{\dagger}
{\bm \sigma}
\bm{\hat{\psi}} ,
\end{eqnarray}
where $\hbar=h/2\pi$ is the Dirac constant, which is the plank constant, $h$, divided by $2\pi$, since each photon brings spin angular momentum of $\hbar$ due to spin 1 character of a photon.
Spin angular momentum operator was derived from the correspondence principle of electromagnetic waves and the rotational symmetry of photons \cite{Saito20a,Saito20c,Saito20d}.
It is straightforward to calculate the expectation values of spin by $|\alpha_{\rm H},\alpha_{\rm V}\rangle$ as 
\begin{eqnarray}
\langle {\bf \hat{S}} \rangle
&=&
\hbar 
\left (
  \begin{array}{c}
    |\alpha_{\rm H}|^2 +   |\alpha_{\rm V}|^2 \\
    \alpha_{\rm H}^{*}\alpha_{\rm V} + \alpha_{\rm V}^{*}\alpha_{\rm H} \\
   -i( \alpha_{\rm H}^{*}\alpha_{\rm V} - \alpha_{\rm V}^{*}\alpha_{\rm H}) 
  \end{array}
\right) \\
&=&
\hbar N
\left (
  \begin{array}{c}
    \cos \gamma \\
    \sin \gamma \cos \delta \\
    \sin \gamma \sin \delta 
  \end{array}
\right),
\end{eqnarray}
which are Stokes parameters \cite{Saito20a} and the polar angle is obtained as $\gamma=2 \alpha$.
Now, we are ready to explain why we could keep the quantum $SU(2)$ states of coherent photons in a pulse during the amplification stage by EDFA.
The important parameters in the $SU(2)$ coherent wavefunction were 2 angles of $\gamma$ and $\delta$.
$\gamma$ is determined by the ratio of $N_{\rm H}$ and $N_{\rm V}$, while $\delta$ is determined by the relative phase among 2 orthogonal oscillations as electromagnetic fields. 
These 2 parameters must not be affected during the amplification process for which we used a polarisation independent fibre.
In the EDFA, the intensity of photons, $I(z)$, is amplified by the gain of $g$ upon the propagation in the fibre as
\begin{eqnarray}
\frac{dI(z)}{dz}=gI(z),
\end{eqnarray}
which obviously gives $I(L)=I_0 {\rm e}^{gL}$, where $I_0$ is the intensity at the input, and $L$ is the length of the gain media.
Accordingly, the number of photons in the pulse also increases to $N=N_0 {\rm e}^{gL}$, where $N_0$ is the number of photons at the input of EDFA.
Consequently, the number of photons were increased, while the polarisation state was not affected.

The process of stimulated emissions is not based on a unitary transformation, such that this is not in contradiction with the non-cloning theorem \cite{Wootters82,Dieks82}.
In fact, the number of photons in the pulse was increased after passing through the EDFA without changing the polarisation state, characterised by $\gamma$ and $\delta$.
For the spin expectation values, the overall factor would be increased due to the increase in $N$, such that the radius of the Poincar\'e sphere was changed.
Mathematically, a sphere with a unit radius is isomorphic to a sphere with a larger radius, since we can assign a bijection function between the points on the 2 spheres with the same $(\gamma,\delta)$, such that 2 spheres are topologically equivalent, regardless of the differences in radius. 
Therefore, the $SU(2)$ state was not affected by the changes in $N$ upon the amplification.
It was also important that we could split the ray into several channels without changing the polarisation state.
In this case, the number of photons in the pulse was decreased, which corresponds to the decrease in the radius.
$N$ is included in the coherent state of $|\alpha_{\rm H},\alpha_{\rm V}\rangle$, and $\langle {\bf \hat{S}} \rangle$ was properly calculated, including the overall factor of $\hbar N$, which is intuitive for reflecting the quantum nature of coherent state, meaning the plank constant effectively increased to be macroscopic.
Thus, physically, the difference in $N$ has several consequences, including a conceptual re-considerations on the capabilities of cloning through stimulated emissions as well as a practical advantage in the improvement of signal-to-noise ratio for polarimetry.
In an $SU(2)$ theory using a Jones vector, the change in $N$ can be easily incorporated through the normalisation of the wavefunction as
\begin{eqnarray}
|N, \gamma, \delta \rangle
=
\sqrt{N}
| \gamma, \delta \rangle
,
\end{eqnarray}
such that the overall norm does not affect the polarisation state.
The increased number of photons contribute to the average spin expectation values of the corresponding spin operator of $\hat{\bf S}=\hbar {\hm \sigma}$ by increasing the radius of the Poincar\'e sphere, whose radius does not have to be normalised to be unity.

\section{Conclusions}

We can accommodate an $SU(2)$ quantum state as a polarisation state in each pulse of coherent photons, emitted from a laser source, such that polarisation-encoded coherent photons are appropriate to be called as a macroscopic qubit.
We showed we can construct an arbitrary single-qubit operation by combining a rotator and a phase-shifter, which we named as a Poincar\'e rotator.
We have experimentally confirmed $SU(2)$ operations of the Poincar\'e rotator, whose signals were analysed by a bespoke polarimeter, observed after the amplifications of the polarisation controlled states.
We have shown that multi-bits of the order of 100 can be distinguishable on the Poincar\'e sphere, and demonstrated to realise topologically the same structures of a C$_{60}$ molecule and coastlines of the earth by photonic states.

\section*{Acknowledgements}
This work is supported by JSPS KAKENHI Grant Number JP 18K19958.
The author would like to express sincere thanks to Prof I. Tomita for continuous discussions and encouragements.


\appendix

\renewcommand{\figurename}{Supplementary Fig}
\setcounter{figure}{0}

\section{Methods}
In this appendix, we explain some details of the measurements.

\subsection{Waveform generations}
We have operated our AWGs at an repetition rate of 10 MHz with a duty cycle of 10 \%.
An example of an pulse as a sine wave, generated by the AWG, is shown in supplementary Fig. 1.
The output signal from the AWG (the red curve) has ripples at the end of the sine curve, which were also amplified after the RF propagation in the electric amplifier (the blue curve).
The ripples were gone after about 10ns from the end of the pulse duration of 10 ns, such that the interference between RF signals could be sufficiently overcome for the duty cycle of 10 \%.

\begin{figure}[h]
\begin{center}
\includegraphics[width=8cm]{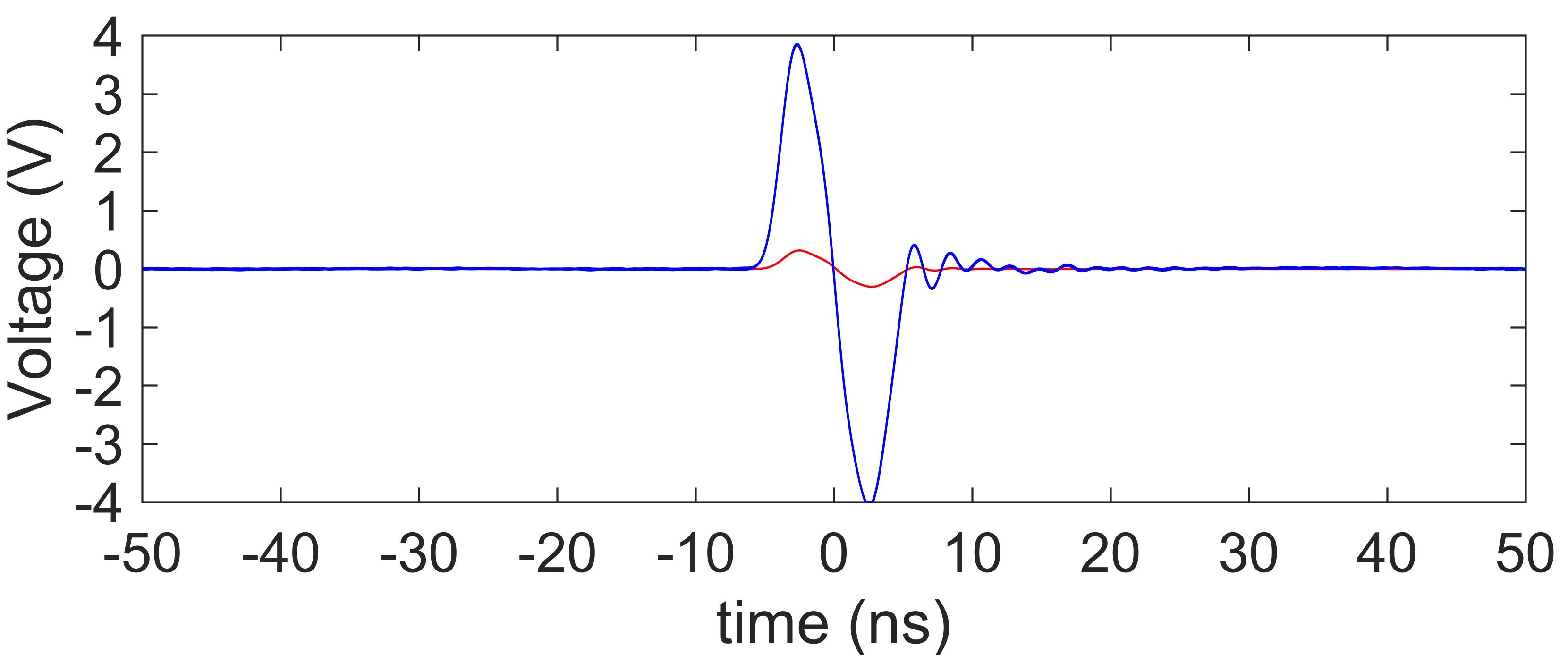}
\caption{
Generation of an electrical signal by an arbitrary-waveform-generator (AWG).
The input waveform is just a simple sine wave with just 1 pulse, operated at 10 MHz with a duty cycle of 10 \%.
The red curve is the measured waveform from an AWG by an oscilloscope at the maximum voltage of 0.3 V.
The blue curve shows the amplified signal out of an electric amplifier, confirming the voltage gain of $\times 12.7$, corresponding to the powr gain of 22.1 dB.
}
\end{center}
\end{figure}

\subsection{Phase-shifter operations}

We have applied the repetition of sine pulses to the phase-shifter, and obtained the outputs from the oscilloscope, as shown in Supplementary Fig. 2.
We have set up the maximum voltage of 0.22 V for the sine pulse from the AWG, which corresponds to the output voltage of $0.22 \times 12.7 = 2.8$ V to the LN modulator.  
This corresponds to the half of the voltage for the phase-shift of $\pi$, $V_{\pi}=0.56$ V for our LN modulator, which is enough to bring the input of the D state to the L and R state, allowing to work as a dynamic QWP.
The maximum voltages of $\pm V_{\pi}$ are required for the modulator in order to scan the entire Poincar\'e sphere.

\begin{figure}[h]
\begin{center}
\includegraphics[width=8cm]{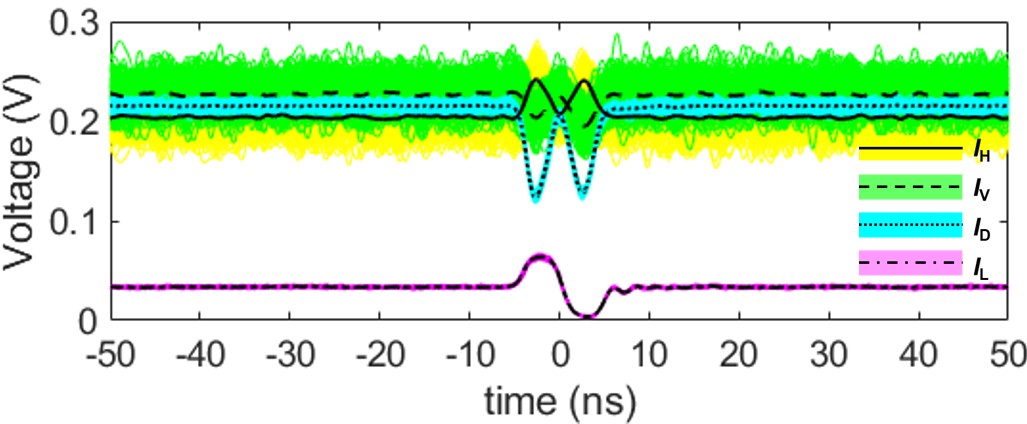}
\caption{
Output signals from a phase-shifter, driven by sine waves with the input of 0.22 V for the AWG.
This corresponds to the application of $\pm V_{\pi}/2 = \pm 2.8$ V to scan from the L state to the R state along parallels in a HV basis.
}
\end{center}
\end{figure}

The variable ranges of the output signals were different among channels, since coupling efficiencies and dark currents were different among PDs.
Therefore, we have normalised the signals based on minimum and maximum values of $I_{\rm H}$, $I_{\rm V}$, $I_{\rm D}$, and $I_{\rm L}$, respectively.
By using the normalised intensities, we have obtained the Stokes parameters.

\subsection{Rotator operations}

\begin{figure}[h]
\begin{center}
\includegraphics[width=8cm]{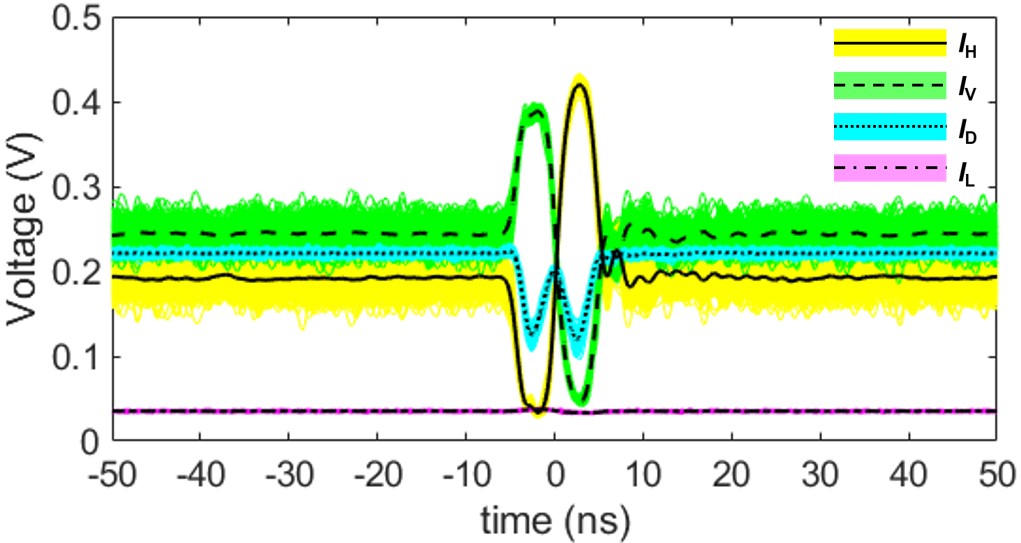}
\caption{
Output signals from a rotator, driven by sine waves with the input of 0.22 V for the AWG.
This corresponds to the application of $\pm V_{\pi}/2 = \pm 2.8$ V to scan from the V state to the H state along meridians in a HV basis.
}
\end{center}
\end{figure}

We have applied the same pulse sequences to the rotator (Supplementary Fig. 3) with those applied to the phase-shifter.
Here, we confirmed changes in $I_{\rm H}$, $I_{\rm V}$, and $I_{\rm D}$, while $I_{\rm L}$ was almost constant, as expected.
This corresponds to the rotation in the $S_1$-$S_2$ plane at $S_3=0$.

\section{Experimental details}
In this appendix, we show some details of the experiments.

\subsection{Multi-bits operations}

\begin{figure}[h]
\begin{center}
\includegraphics[width=8cm]{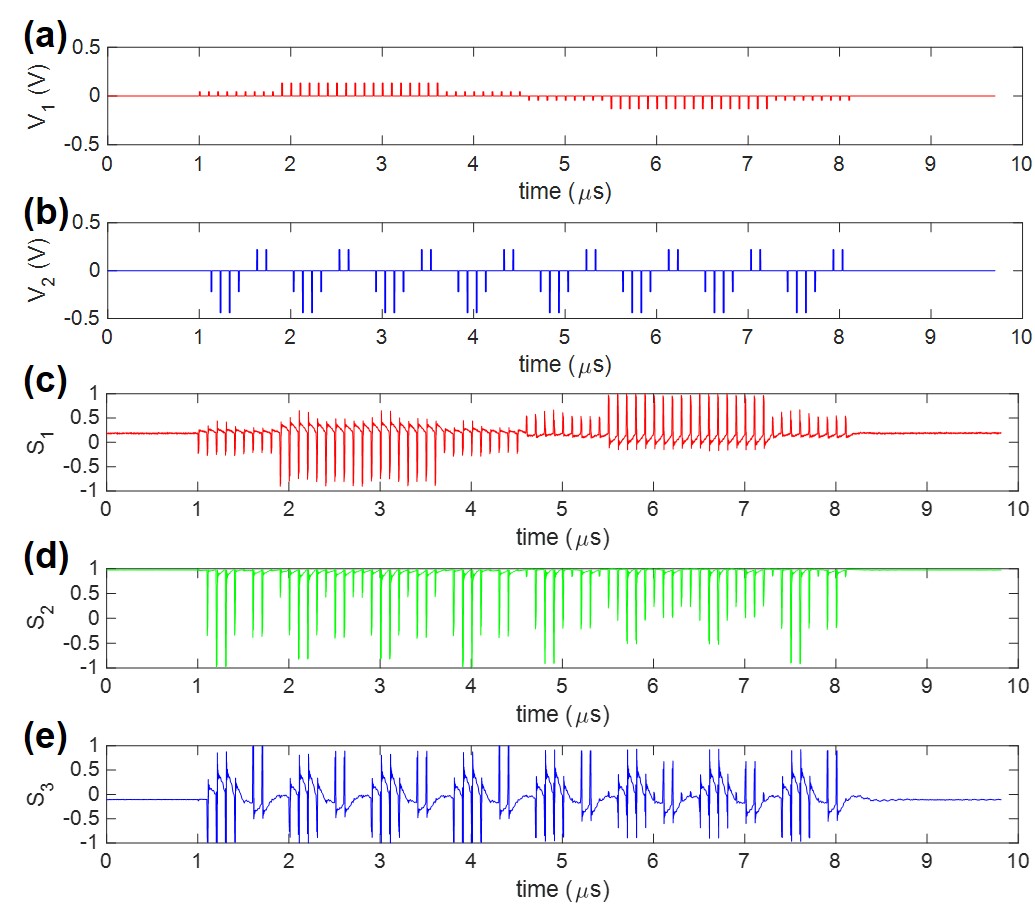}
\caption{$4 \times 4=16$ bits on the Poincar\'e sphere. 
Input waveforms of AWGs for (a) a rotator and (b) a phase-shifter. 
Output waveforms of Stokes parameters, (c) $S_1$, (d) $S_2$, and (e) $S_3$.
}
\end{center}
\end{figure}

The input waveforms for AWGs are shown in Supplementary Figs. 4 (a) and (b). 
The maximum values of 0.22 V and 0.44 V were applied to $V_1$ and $V_2$, respectively.
The application of $V_1$ to the rotator induced the rotation along the $S_3$ axis, which  changed $S_1$ and $S_2$.
In the present experimental set-up (Fig. 3), the application of $V_1>0$ corresponded to the anti-clock-wise rotation (the positive rotation for $\gamma$), which reduced the values of $S_1$ located at the input state of the D state.
In fact, we confirmed that the signs of $S_1$ values were opposite to those of $V_1$ values  (Supplementary Fig. 4 (c) ).
Similarly, the application of $V_2>0$ corresponded to the anti-clock-wise rotation (the positive rotation for $\delta$) along the $S_1$ axis, which increased the values of $S_3$ from the D state, such that the signs of $S_3$ values were the same as those of $V_2$ values (Supplementary Fig. 4 (e) ).

The input waveforms for AWGs and output Stokes parameters for {$8 \times 8=64$ bits and  {$10 \times 10=100$ bits are shown in Supplementary Figs. 5 and 6, respectively.

\begin{figure}[h]
\begin{center}
\includegraphics[width=8cm]{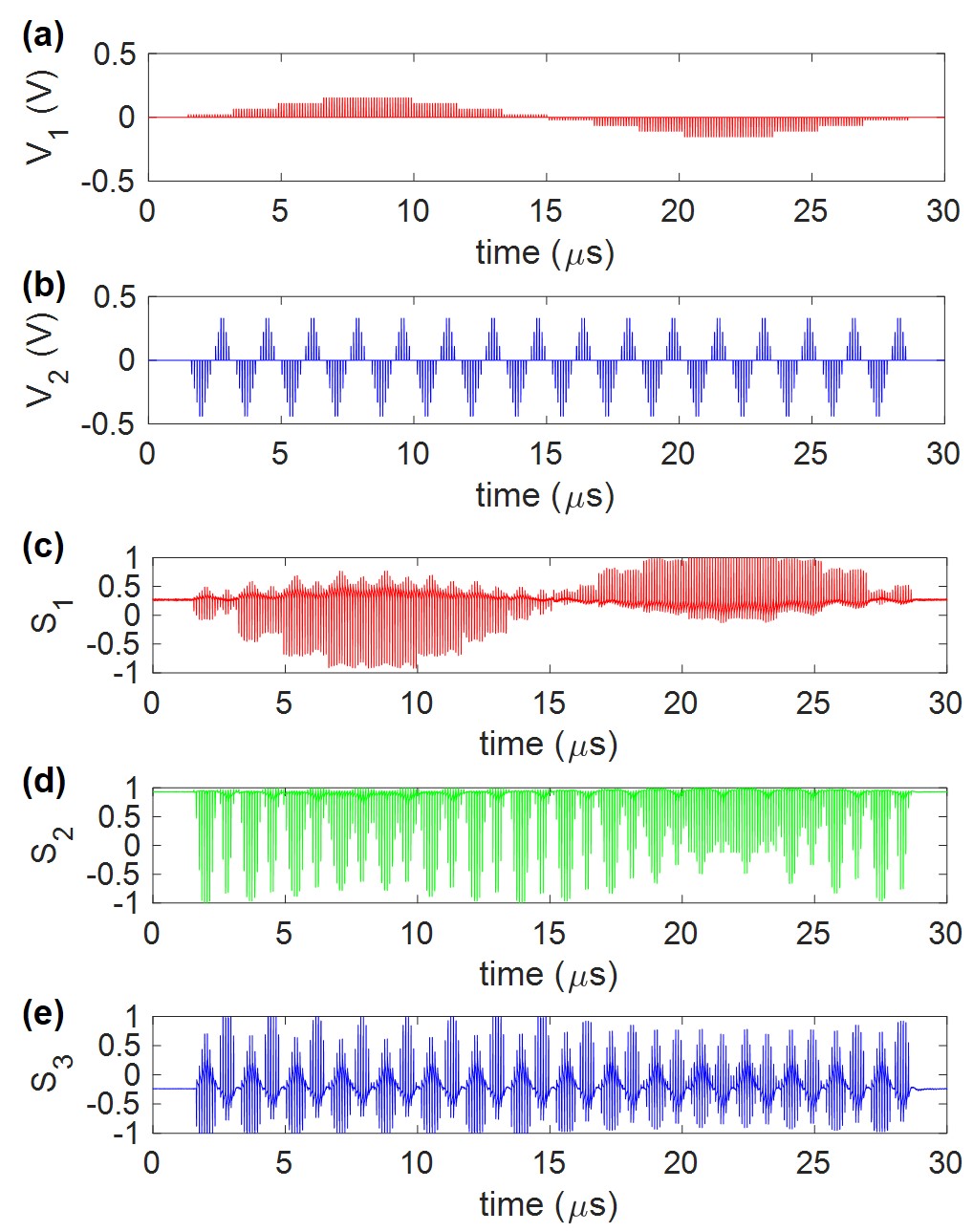}
\caption{$8 \times 8=64$ bits on the Poincar\'e sphere. 
Input waveforms of AWGs for (a) a rotator and (b) a phase-shifter. 
Output waveforms of Stokes parameters, (c) $S_1$, (d) $S_2$, and (e) $S_3$.
}
\end{center}
\end{figure}

\begin{figure}[h]
\begin{center}
\includegraphics[width=8cm]{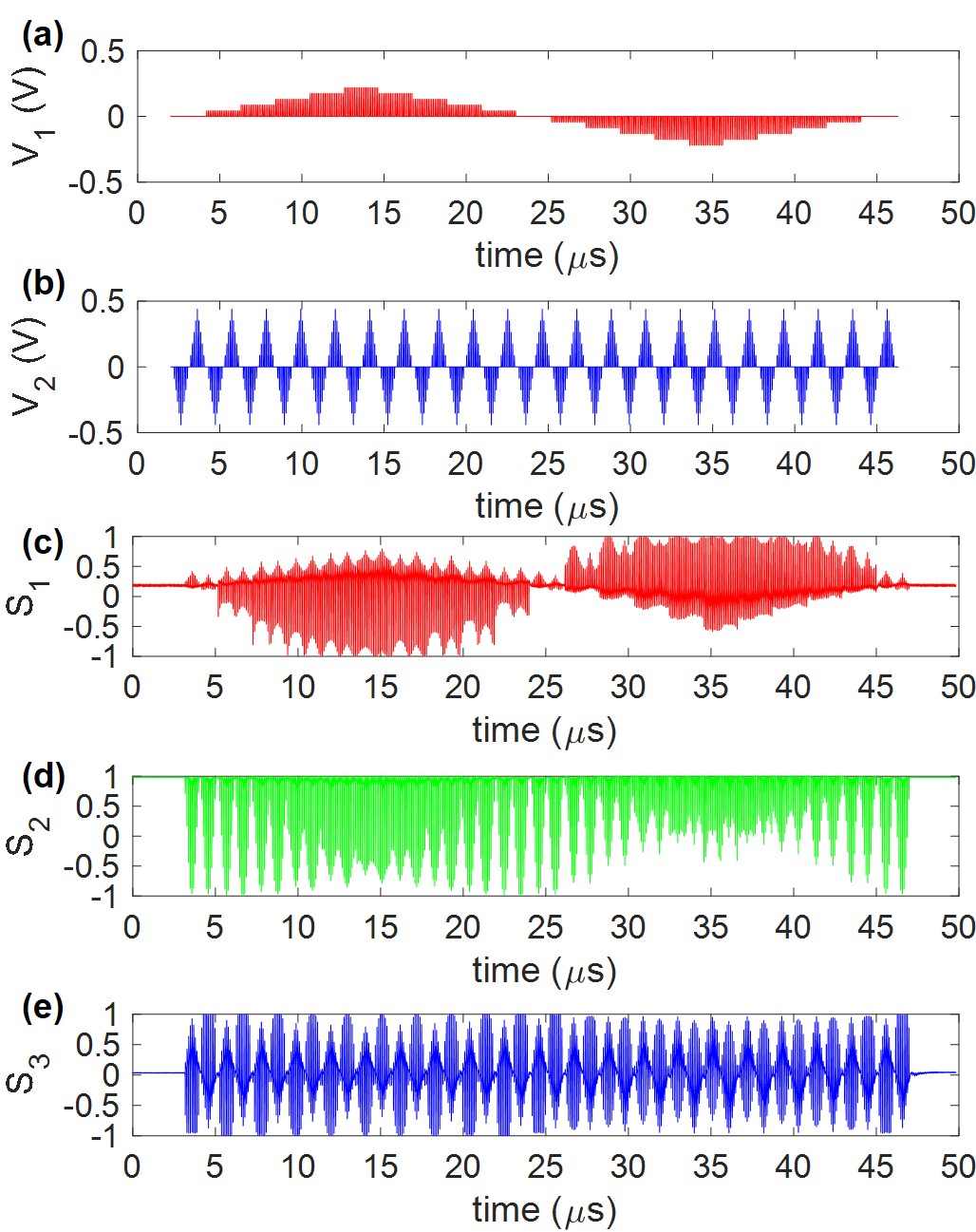}
\caption{$10 \times 10=100$ bits on the Poincar\'e sphere. 
Input waveforms of AWGs for (a) a rotator and (b) a phase-shifter. 
Output waveforms of Stokes parameters, (c) $S_1$, (d) $S_2$, and (e) $S_3$.
}

\end{center}
\end{figure}

\subsection{Drawing for C$_{60}$ and the earth}

\begin{figure}[h]
\begin{center}
\includegraphics[width=8cm]{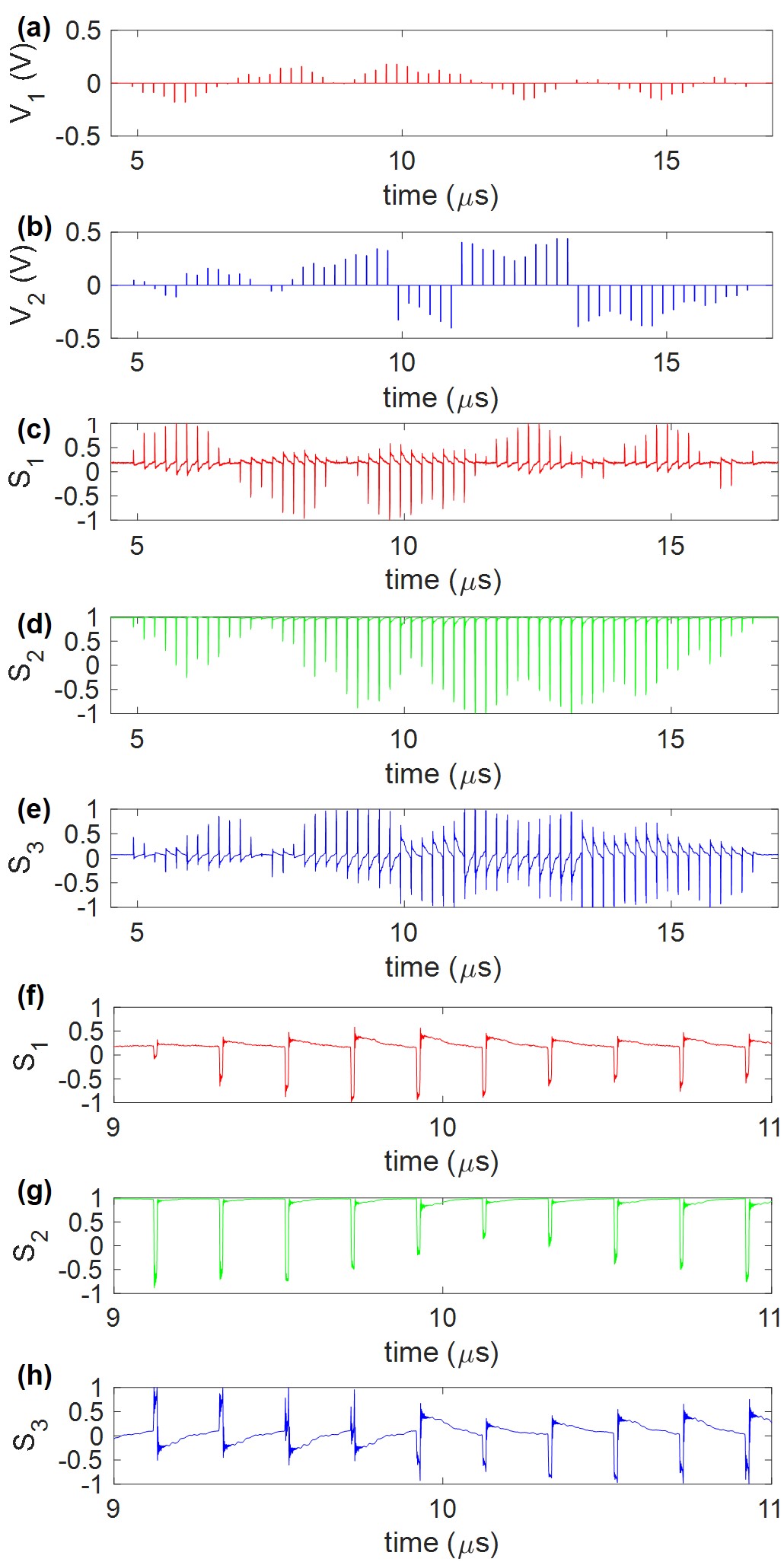}
\caption{
Fullerene C$_{60}$ on the Poincar\'e sphere by a left Hamilton path.
Input waveforms of AWGs for (a) a rotator and (b) a phase-shifter. 
Output waveforms of Stokes parameters, (c) $S_1$, (d) $S_2$, and (e) $S_3$, and magnified pulse shapes of (f) $S_1$, (g) $S_2$, and (h) $S_3$. were observed
}
\end{center}
\end{figure}

The left Hamilton path was calculated for C$_{60}$ and converted as the inputs for a rotator and a phase-shifter (Supplementary Figs. 7 (a) and (b)).
60 sets of voltages $(V_1,V_2)$ corresponds to the vertex coordinates $(\gamma,\delta)$ on the Poincar\'e sphere.
As the definition of the Hamilton path, each vertex was visited only once during the modulations. 
We obtained the full trajectories of Stokes parameters (Supplementary Figs. 7 (c)-(e)), modulated for the input of the D state.
As shown in the magnified outputs in Supplementary Figs. 7 (f)-(h), the output signals were distorted after switching-off the pulses due to the ripples of our return-to-zero signals from AWGs.
Therefore, we took the longer period of 200 ns to avoid interferences between pulses.

\begin{figure}[h]
\begin{center}
\includegraphics[width=8cm]{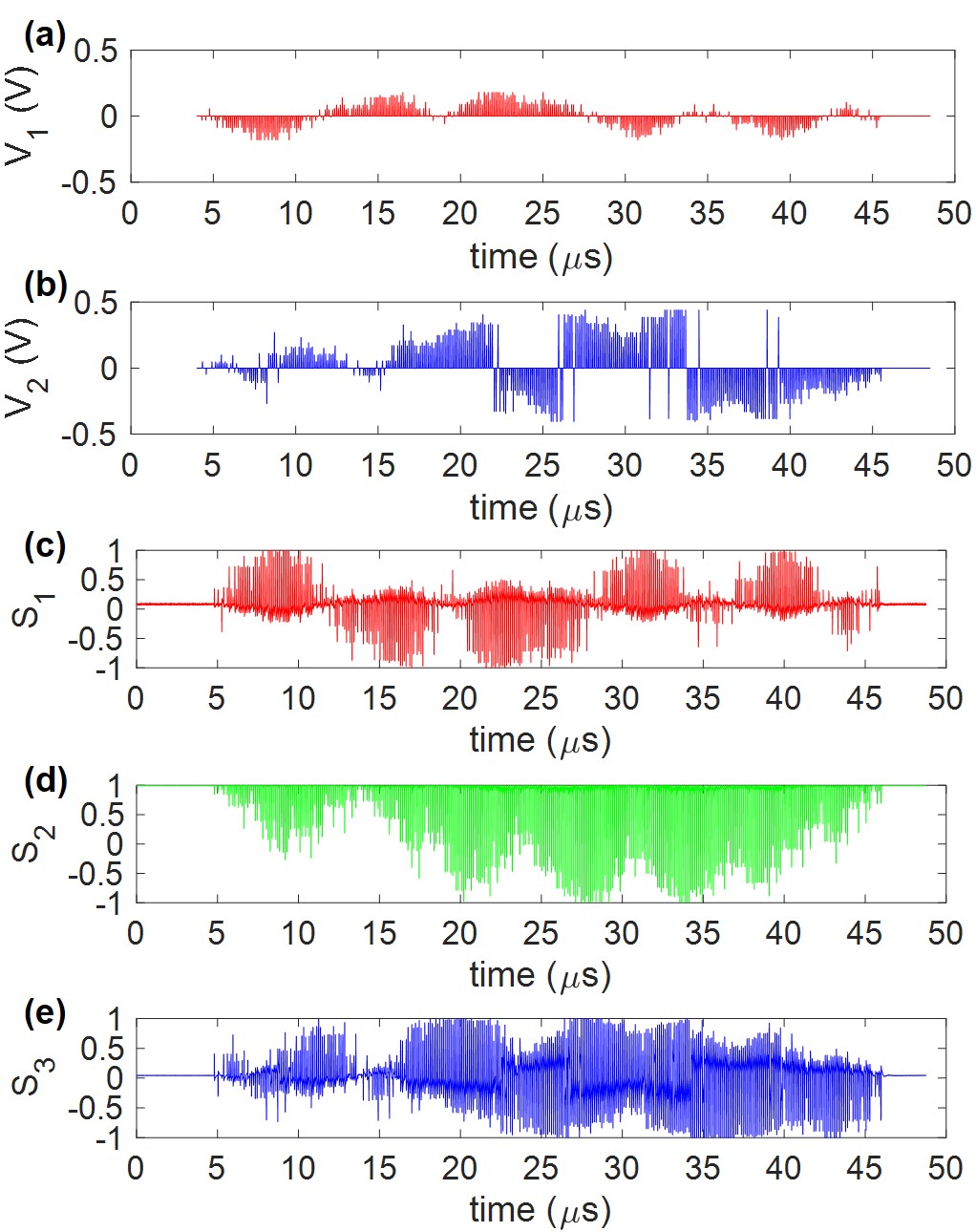}
\caption{
Fullerene C$_{60}$ on the Poincar\'e sphere to realise all bonding.
Input waveforms of AWGs for (a) a rotator and (b) a phase-shifter, were calculated by using a left Hamilton path adding 3 nearest neighbouring vertexes to visit. 
Output waveforms of Stokes parameters, are shown in (c) $S_1$, (d) $S_2$, and (e) $S_3$.
}
\end{center}
\end{figure}

We have also tried to realise all bonds of C$_{60}$ on the Poincar\'e sphere, as shown in supplementary Fig. 8.
Here, bonds are represented by the nearest neighbouring vertexes on the the Poincar\'e sphere.
We used the left Hamiltonian path for C$_{60}$ to visit a vertex, and at each vertex, we have set to visit 3 nearest neighbouring vertexes before going to the next vertex in the left Hamiltonian path.
In this case, each vertex was visited 3 times during the scan.

\begin{figure}[h]
\begin{center}
\includegraphics[width=8cm]{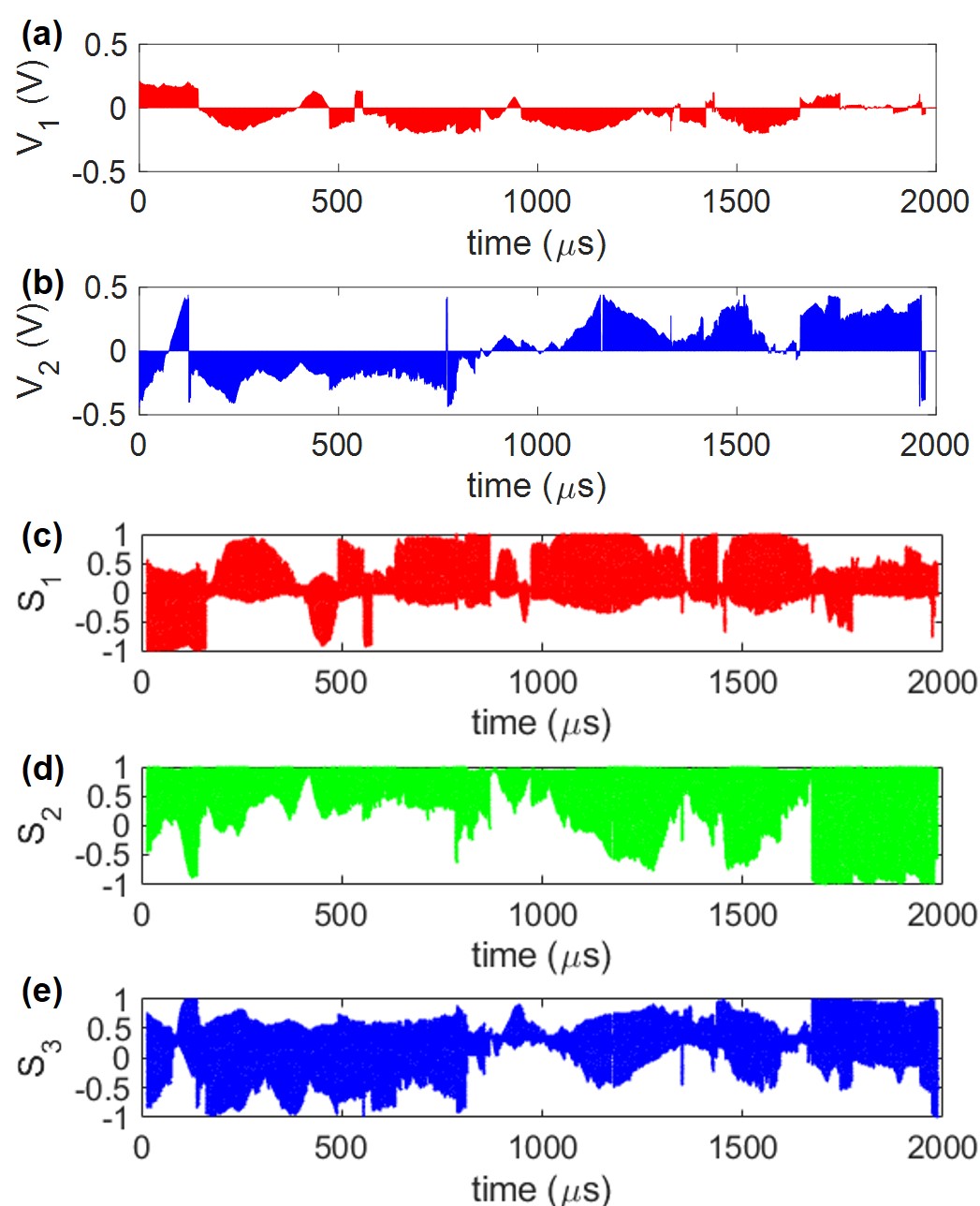}
\caption{
Coastlines of the earth on the Poincar\'e sphere as polarisation states of pulses.
Input waveforms of AWGs for (a) a rotator and (b) a phase-shifter, were prepared by converting geometrical data to the voltages for modulators. 
Output waveforms of Stokes parameters, are shown in (c) $S_1$, (d) $S_2$, and (e) $S_3$.
}
\end{center}
\end{figure}

\clearpage
\bibliography{ActivePR}

\end{document}